\documentclass{article}
\usepackage{graphicx} 

\usepackage{authblk}     
\usepackage{geometry}    
\usepackage{graphicx}    
\usepackage{amsmath, amssymb}
\usepackage{natbib}      
\usepackage{hyperref}    
\usepackage{indentfirst}
\usepackage{booktabs}
\numberwithin{equation}{section}
\usepackage{makecell}

\title{On the role of back-propagating pressure suppression in enhancing the pressure-gain performance of quasi-2D rotating detonation engines}

\author[1]{Tonghui Wang}
\author[1]{Guoqing Zhang\thanks{Corresponding author: \texttt{zhanggq@bit.edu.cn}}}
\author[2]{Haocheng Wen\thanks{Corresponding author: \texttt{haochengwenson@126.com}}}

\affil[1]{School of Aerospace Engineering, Beijing Institute of Technology, Beijing 100081, China}
\affil[2]{School of Aerospace Engineering, Tsinghua University, Beijing 100084, China}

\date{} 

\begin{document}

\maketitle

\paragraph{Abstract:} The total pressure gain (PG) characteristics of the quasi-2D rotating detonation engine (RDE) are numerically investigated in this study, based on an abstract check valve model and the quasi-1D assumption. The influence of back-propagating pressure suppression on PG and its underlying mechanism are examined. An abstract check valve model is established to simulate various flow channel configurations, with backflow check strength $\alpha_b$ defined, where a larger $\alpha_b$ corresponds to a stronger backflow blocking effect. The quasi-1D assumption is applied along the axial direction to simplify the radial features of the annular RDE. The quasi-2D governing equations for RDE flow are derived. Simulations are conducted for varying expansion ratios $A_e$ and values of $\alpha_b$. The results indicate that increasing $\alpha_b$ effectively suppresses back-propagating pressure and slightly improves PG; however, it cannot fully eliminate the back-propagating pressure, as the check valve itself introduces flow disturbances. Increasing $A_e$ also suppresses back-propagating pressure but significantly reduces PG. Achieving positive PG requires reducing $A_e$ below a critical value. However, this reduction is limited by $\alpha_b$; further reduction in $A_e$ leads to forward propagation of back-propagating pressure to the engine inlet, resulting in inlet blocking. Therefore, a sufficiently large $\alpha_b$ is essential for the required reduction in $A_e$. The key aerodynamic challenge for achieving positive PG lies in optimizing flow channels to suppress back-propagating pressure efficiently. Finally, a general PG criterion is proposed by normalizing the quasi-2D RDE with stoichiometric hydrogen/air mixtures. This study provides theoretical guidance for enhancing PG in RDEs.


\begin{table}[h!] 
  \centering 
  \caption{Nomenclature}
  \label{tab:nomenclature}
  \setlength{\tabcolsep}{10pt} 
  \begin{tabular}{l p{10cm}} 
    $A$          & Area ratio \\
    $p$          & Pressure \\
    $T$          & Temperature \\
    $\rho$       & Density \\
    $u$          & Spanwise velocity \\
    $v$          & Axial velocity \\
    $A_e$        & Expansion ratio \\
    $\alpha_b$   & Backflow check strength \\
    $\alpha_f$   & Forward flow check strength \\
    $M$          & Mach number \\
    $\theta$     & Oblique shock wave inclination angle \\
    $\eta$       & Total pressure recovery coefficient \\[6pt] 
    
    \multicolumn{2}{l}{\textbf{Superscript}} \\[3pt]
    $\overline{\phantom{x}}$ & Averaged parameters \\[6pt]
    
    \multicolumn{2}{l}{\textbf{Subscript}} \\[3pt]
    $lb$         & Lower boundary of the check valve model \\
    $ub$         & Upper boundary of the check valve model \\
    $21$         & Station where the reverse oblique shock wave deflects \\
    $22$         & Station where the reverse oblique shock wave intersects with itself \\
    $31$         & Outlet of the expansion section \\
    $32$         & Inlet of the combustor \\ 
    $4$          & Outlet of the engine \\
    $t$          & Total parameters \\
    $cr$         & Critical parameters \\
  \end{tabular}
\end{table}

\clearpage
\section{Introduction}
\label{sec1}

Unlike deflagration, where the flame propagates at subsonic speed, continuous detonation exhibits advantages of fast heat release and supersonic flame, by tightly coupling the leading shock wave with the chemical reaction zone \citep{Fickett01}. As a new type of detonation propulsion device, the rotating detonation engine (RDE) has outstanding advantages in propulsion performance due to its high energy conversion efficiency. In recent years, RDE has received extensive attention, with numerous numerical simulations \citep{Xia02,Zhao03,Meng04,Prakash05} and experimental studies \citep{Fan06,Xia07,Kindracki08,Zhou09} conducted to explore the mode transition mechanism of detonation waves and analyze the key factors influencing engine performance.

As a key indicator for assessing the performance of RDEs compared to conventional engines, the total pressure gain (PG) has long been a focus of research. However, controversy remains regarding the evaluation of RDE's PG and whether RDEs can achieve positive PG.

Some studies evaluating PG through numerical simulations argue that RDE achieves positive PG under certain conditions. For instance, \citet{Kaemming10} employs the equivalent available pressure (EAP) method to calculate PG based on RDE simulations. The results indicate that PG increases as the combustor inlet injection area increases and the exit throat cross-sectional area decreases. Furthermore, when the inlet injection area is sufficiently large and the exit throat area is sufficiently small, RDE achieves positive PG. Other numerical studies \citep{Wang36,Liu37,Qin39} also suggest that modifying injection methods is effective for achieving positive PG. \citet{Sheng11} conducts numerical simulations to investigate the effect of multiple detonation waves on RDE performance, deriving PG via the mass-averaged total pressure (MAP) method. Their results show that RDE achieves positive PG under various detonation wave numbers, with the PG for the single-wave mode being higher than under other conditions. \citet{Zhang12} examines the variation of the average total pressure at the RDE exit with the injection total temperature through numerical simulations. They find that the average total pressure decreases as the injection temperature increases, and that when the total temperature is below 600K, RDE consistently achieves positive PG. \citet{Wang38} finds that smaller carbon particle diameters help maintain positive PG, based on outlet average total pressure numerical results.

However, most experimental studies to date fail to achieve positive PG in RDE. For instance, \citet{Bach13} use Kiel probes to directly measure the stagnation pressure of high-enthalpy exhaust flows and investigate the effect of the rotating detonation combustor (RDC) geometric configuration on PG. The results show that the maximum achievable PG is $-$8\%. They also compile the PG data from other experimental studies \citep{Brophy14,TenEyck15,Walters16,Rankin17,Frolov18,Bach19} and find that none of these experiments achieve positive PG. \citet{Kang40} measure the exit total pressure of the rotating detonation afterburner (RDAB) via the Mach number corrected static pressure (MCSP) method. They find that when the nozzle exit area ratio is relatively high, adopting the combined injection scheme can effectively enhance the detonation wave intensity but fails to improve the total pressure recovery coefficient. As a result, positive PG cannot be achieved. Other experimental investigations, covering different fuel types and operating pressures \citep{Feleo41,Hayashi42,Plaehn43}, also support the conclusion that PG remains negative in RDE under current test conditions.

Clearly, there are differences between the conclusions of numerical and experimental studies. We believe the reason for these differences lies in the different PG evaluation methods. More importantly, however, rotating detonation waves induce reverse shock waves in practice, with back-propagating pressure exerting a significant impact on the engine’s performance. The effect of this phenomenon on PG has not been accurately assessed yet. Several studies have already analyzed this issue.

\citet{Wu20} investigate the axial distribution of the average total pressure in the air-breathing RDC under different inlet Mach numbers via numerical simulations. The results show that, compared to the throat state, the total pressure of the gas decreases significantly by 25\% due to the effects of upstream oblique shock waves and the geometric configuration, which ultimately prevents the RDE from achieving positive PG. \citet{Chen21} investigate the axial distribution of MAP in the rotating detonation ramjet engine (RDRE) under different combustor configurations via numerical simulations. They find that, for the same length, the total pressure loss in the divergent combustor is 7.25\% less than that in the isometric combustor. However, due to the total pressure loss caused by the leading shock waves in the isolator and divergent section, RDREs fail to achieve positive PG. \citet{Wen22} propose a PG evaluation model through theoretical analysis and the MAP method, based on the inlet Mach number and the anti-backflow capability of the rotating detonation system. The results indicate that positive PG of the system cannot be guaranteed at all times. The key to achieving positive PG is suppressing reverse oblique shock waves (ROSW) using the ideal check valve and adjusting the inlet Mach number to be close to 1. \citet{Jiang44} and \citet{Cao45} numerically study the axial MAP of the two-phase kerosene/air RDE, finding that the upstream oblique shock wave and the normal shock in the divergent section cause large total pressure loss, resulting in negative PG. The above studies indicate that back-propagating pressure exerts a negative impact on the pressure-gain performance of RDEs.

To suppress the back-propagating pressure, a series of relevant studies focus on aspects such as injection structure optimization, isolator structure design, and one-way flow control structure optimization. \citet{Schwer23,Schwer24} find that a relatively large injection area ratio leads to strong pressure disturbances in the plenum via numerical simulations. To reduce such disturbances, they further investigate the effect of injection structure geometries (including Cavity-slot, Nozzle-slot, Diode-slot, and slot inclination angle). The results show that none of these structures effectively suppress the pressure feedback. \citet{Frolov25} design an upstream isolator with a maximum width of $\Delta$. Numerical simulations show that increasing $\Delta$ effectively prevents the upstream propagation of pressure disturbances. \citet{Ji26} conduct numerical simulations of internal flow to investigate the propagation of pressure disturbances. Based on this, they design a cylindrical isolator structure with multiple rows of wedge obstacles arranged on its inner wall and propose preliminary optimization design criteria for the isolator. The results show that increasing the number, dimensionless height, width, and spacing of the obstacles helps reduce feedback pressure perturbation, while increasing the inclination angle has the opposite effect. \citet{Yang27,Yang28} propose an optimized Tesla valve structure with a bypass manifold. Experimental studies show that, compared with the convergent-divergent intake configuration, Tesla valve structure without a bypass manifold, and the unoptimized Tesla valve structure, this intake configuration exhibits the lowest pressure feedback percentage (about 6.5\%).

Although the above studies clearly clarify two aspects: first, that back-propagating pressure is one of the reasons why RDE struggles to achieve positive PG; second, that a series of studies on flow channel configuration optimization have been conducted to suppress the back-propagating pressure. There remains a lack of systematic and in-depth research on how different suppression levels of back-propagating pressure influence PG. Meanwhile, no clear conclusion has been reached regarding whether positive PG can be achieved in RDE and the corresponding conditions for it. In this study, we propose a check valve model capable of suppressing back-propagating pressure to simulate the specific flow channel configuration. Axially, we simplify the radial features of the annular RDE using the quasi-1D assumption, and conduct various numerical simulations of the quasi-2D RDE with the expansion section. The variation of PG with the degree of back-propagating pressure suppression is explored, and the mechanism underlying the failure to achieve positive PG is analyzed.

The following is divided into three sections. In section~\ref{sec2}, the governing equations of the check valve model are constructed. The existence of the model's physical solution is then analyzed, and its feasibility for simulating the specific flow channel configuration is verified. In section~\ref{sec3}, the numerical method and physical model are elaborated. Under the quasi-1D assumption, the governing equations for the quasi-2D RDE flow are established, and the specific process of applying the check valve model to solve boundary conditions is explained. In section~\ref{sec4}, the effects of back-propagating pressure suppression on the internal flow field structure and pressure-gain performance of the quasi-2D RDE are discussed, and the conditions for achieving positive PG are clarified.

\section{Check valve model}
\label{sec2}

\subsection{Basic structure}
\label{sec2.1}

\begin{figure}
  \centerline{\includegraphics[width=\textwidth]{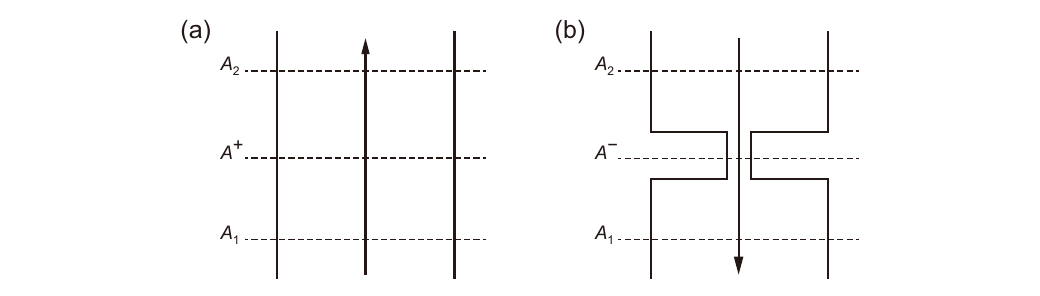}}
  \caption{Schematic diagram of the check valve model \citep{Wen22}: (a) forward flow schematic, (b) backflow schematic.}
\label{fig1}
\end{figure}

Although a large number of studies focus on optimizing flow channels to suppress back-propagating pressure \citep{Schwer23,Schwer24,Frolov25,Ji26,Yang27,Yang28}, it remains difficult to systematically discuss the effects of diverse flow channel configurations. For this reason, this study proposes an abstract check valve model, as shown in figure~\ref{fig1}, to simulate these configurations with different flow characteristics. The model is designed to balance normal fuel injection with the suppression of back-propagating pressure from the combustor. The core design concept involves differentially adjusting the channel areas for each flow direction, ensuring the normal progression of the forward flow while blocking the backflow.

Assuming that the inlet and outlet cross-sectional areas of the check valve model are equal, i.e., $A_1 / A_2 = 1$, when the gas undergoes forward flow (from position 1 to position 2) and backflow, the corresponding throat cross-sectional areas of the check valve model are $A^+$ and $A^-$, respectively. Here, $A^+ / A_1 = 1$, indicating that during forward flow, the channel is fully open, so the check valve model does not exert any force on the gas. However, $A^- / A_2 < 1$, meaning that during backflow, the channel contracts, and the check valve model generates a block to the gas.

\subsection{Governing Equations}
\label{sec2.2}

\begin{figure}
  \centerline{\includegraphics[width=\textwidth]{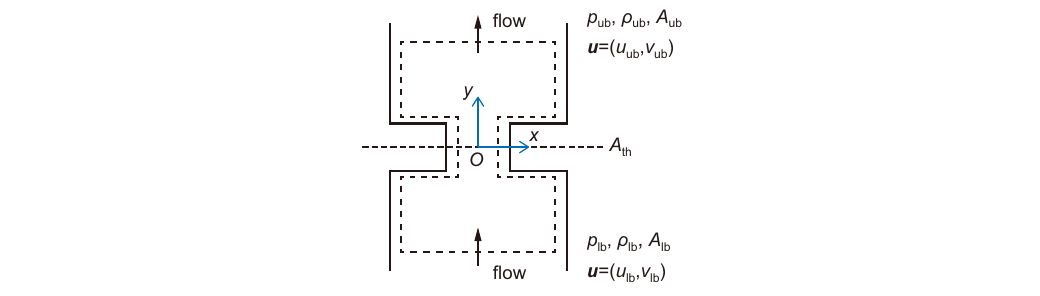}}
  \caption{Control volume of the check valve model.}
\label{fig2}
\end{figure}

The governing equations of the check valve model are established using the control volume method, as shown in figure~\ref{fig2}. The subscripts ub and lb are used to denote the flow parameters at the upper and lower boundaries of the check valve model, respectively, and the throat cross-sectional area is denoted as $A_{th}$.

It is assumed that the flow is steady, inviscid, and adiabatic, with uniform flow across the inlet and outlet cross-sections. Additionally, the gas is considered thermally perfect, meaning its specific heats $C_p$ and $C_v$ are constant, and it satisfies the ideal gas equation of state, $p = \rho RT$. Based on the work of \citet{Wen30}, the governing equations for the check valve model are established as follows:

\begin{equation}
  \rho_{lb} v_{lb}=\rho_{ub} v_{ub}
  \label{a2.1}
\end{equation}
\begin{equation}
  p_{lb} A_{lb}+\rho_{lb} v_{lb}^2 A_{lb}+F=p_{ub} A_{ub}+\rho_{ub} v_{ub}^2 A_{ub}
  \label{a2.2}
\end{equation}
\begin{equation}
  u_{lb}=u_{ub}
  \label{a2.3}
\end{equation}
\begin{equation}
  \frac{\gamma}{\gamma-1} \frac{p_{lb}}{\rho_{lb}} + \frac{1}{2} v_{lb}^2 
  = \frac{\gamma}{\gamma-1} \frac{p_{ub}}{\rho_{ub}} + \frac{1}{2} v_{ub}^2
  \label{a2.4}
\end{equation}

Where $F$ denotes the force exerted by the check valve model on the control volume. The drag coefficient $C_F$ ($C_F < 0$) is used to calculate it as follows:

\begin{equation}
  F=\frac{C_F}{2} \rho_{lb} v_{lb}^2 \left(A_{lb}-A_{th} \right)
  \label{a2.5}
\end{equation}

Define the check strength $\alpha$, whose expression is given as follows:

\begin{equation}
  \alpha=\frac{C_F}{2} \left(\frac{A_{th}}{A_{lb}} -1\right)-1
  \label{a2.6}
\end{equation}

Substituting equations (\ref{a2.5}) and (\ref{a2.6}) into equation (\ref{a2.2}) yields:

\begin{equation}
  p_{lb}-\rho_{lb} v_{lb}^2 \alpha=p_{ub}+\rho_{ub} v_{ub}^2
  \label{a2.7}
\end{equation}

The Rayleigh line and the Hugoniot curve of the check valve model can be obtained by performing algebraic operations on equations (\ref{a2.1}), (\ref{a2.4}), and (\ref{a2.7}).

The density ratio $x$ and pressure ratio $y$ are defined as follows:

\begin{equation*}
  x=\frac{\rho_{lb}}{\rho_{ub}} ,y=\frac{p_{ub}}{p_{lb}} 
\end{equation*}

The expression for the Rayleigh line is given as follows:

\begin{equation}
  y=-\gamma M_{lb}^2 x-\alpha \gamma M_{lb}^2+1
  \label{a2.8}
\end{equation}

The expression for the Hugoniot curve is given as follows:

\begin{equation}
  y=\frac{-\frac{1}{2} x^2+\frac{\gamma}{\gamma-1} x+\frac{1}{2}+\alpha \frac{\gamma}{\gamma-1}}{\frac{\gamma+1}{2(\gamma-1)}  x^2+\alpha \frac{\gamma}{\gamma-1} x+\frac{1}{2}}
  \label{a2.9}
\end{equation}

\subsection{Physical solution existence}
\label{sec2.3}

To ensure the solutions of the check valve model (the intersections of Rayleigh lines and Hugoniot curves) are physical, the following two constraint conditions \citep{Wen30} must be satisfied:

(1) The Mach number $M_{lb}$ at the lower boundary of the check valve model is a real number, and its expression is given as follows:

\begin{equation}
  M_{lb}^2=\frac{1}{\gamma}  \frac{1-y}{\alpha+x}>0
  \label{a2.10}
\end{equation}

(2) Since there is no heat exchange process when the gas flows through the check valve model, the entropy change of the gas must be greater than 0. The formula for calculating entropy is given as follows:

\begin{equation}
  \Delta s = \frac{R}{\gamma - 1} \ln\left(y x^\gamma\right) > 0
  \label{a2.11}
\end{equation}

It is clear from equations (\ref{a2.8}) to (\ref{a2.11}) that the existence conditions of physical solutions are directly related to the parameter $\alpha$. Therefore, the check strength for forward flow is defined as $\alpha_f$, and that for backflow is defined as $\alpha_b$. Systematic analyses of how solution existence relates to the value ranges of $\alpha_f$ and $\alpha_b$ are presented in appendix~\ref{appA} and appendix~\ref{appB}, respectively. The results show that when the gas flows forward through the check valve model, the flow parameters at the upper and lower boundaries of the model are completely consistent under the isentropic assumption. When the gas flows back through the check valve model, further organizing the statistics in table~\ref{tab4} of appendix~\ref{appB} leads to the following conclusions:

(1) When $\alpha_b \in (-1,\alpha_1)$, there are two cases in which $M_{lb}$ is real: first, when $x \in (x_{f1},x_{f2})$, in which case $y > 1$, corresponding to the flow from the low-pressure region to the high-pressure region; second, when $x \in (1,x_{g2})$, in which case $0 < y < 1$, corresponding to the flow from the high-pressure region to the low-pressure region.

(2) When $\alpha_b \in (\alpha_1,\infty)$, $M_{lb}$ is real only when $x \in (1,x_{g2})$, in which case $0 < y < 1$, and the gas exhibits the characteristic of flowing from the high-pressure region to the low-pressure region.

The calculation method for the PG of the check valve model is defined as follows:

\begin{equation}
  PG = \frac{p_{ub}}{p_{lb}} 
  \frac{\left(1 + \frac{\gamma-1}{2} M_{ub}^2 \right)^{\frac{\gamma}{\gamma-1}}}
       {\left(1 + \frac{\gamma-1}{2} M_{lb}^2 \right)^{\frac{\gamma}{\gamma-1}}} - 1 
  = y \left( \frac{1 + \frac{\gamma-1}{2\gamma} \frac{x}{y} \frac{1-y}{x+\alpha_b}}
                 {1 + \frac{\gamma-1}{2\gamma} \frac{1-y}{x+\alpha_b}} 
       \right)^{\frac{\gamma}{\gamma-1}} - 1
  \label{a2.12}
\end{equation}

\begin{figure}
  \centerline{\includegraphics[width=\textwidth]{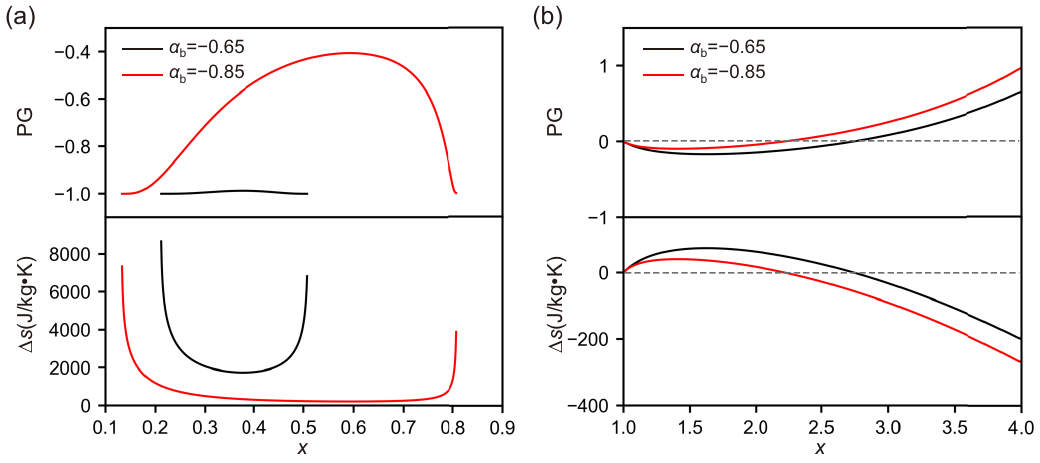}}
  \caption{Variations of entropy change $\Delta s$ and PG with respect to the density ratio $x$ under the condition where the backflow check strength $\alpha_b \in (-1, \alpha_1)$: (a) for $x \in (x_{f1}, x_{f2})$, and (b) for $x \in (1, x_{g2})$.}
\label{fig3}
\end{figure}

\begin{figure}
  \centerline{\includegraphics[width=\textwidth]{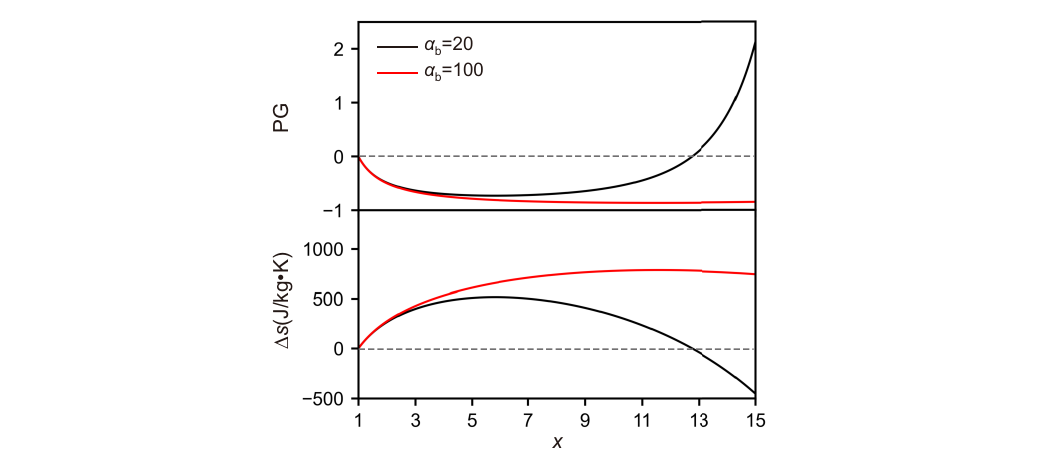}}
  \caption{Variations of entropy change $\Delta s$ and PG with respect to the density ratio $x$ under the condition where the backflow check strength $\alpha_b \in (\alpha_1, \infty)$ for $x \in (1, x_{g2})$.}
\label{fig4}
\end{figure}

Let the specific heat ratio $\gamma = 1.24$, then the interval division point $\alpha_1$ in appendix~\ref{appB} is $-0.59$. To investigate the influence of the backflow check strength $\alpha_b$ on PG, four typical values of $\alpha_b$ are selected for a comparative study across intervals: $\alpha_b \in (-1,\alpha_1)$ with values $-0.65$ and $-0.85$, and $\alpha_b \in (\alpha_1,\infty)$ with values 20 and 100. For the cases where $\alpha_b = -0.65$ and $-0.85$, the variations of entropy change $\Delta s$ and PG with respect to the density ratio $x$ are shown in figure~\ref{fig3}. The variations of the same parameters for the cases where $\alpha_b = 20$ and 100 are shown in figure~\ref{fig4}. Analysis of the results in figures~\ref{fig3} and~\ref{fig4} reveals that, under the condition of satisfying equation (\ref{a2.11}), an increase in check strength $\alpha_b$ at the same density ratio $x$ leads to a reduction in PG, indicating that the check valve model exerts a stronger blocking effect on the backflow. To ensure that the gas always flows from the high-pressure region to the low-pressure region during the backflow, the value of $\alpha_b$ in subsequent studies will be strictly limited to the range above $\alpha_1$.


\subsection{Model feasibility verification}
\label{sec2.4}

\begin{figure}
  \centerline{\includegraphics[width=\textwidth]{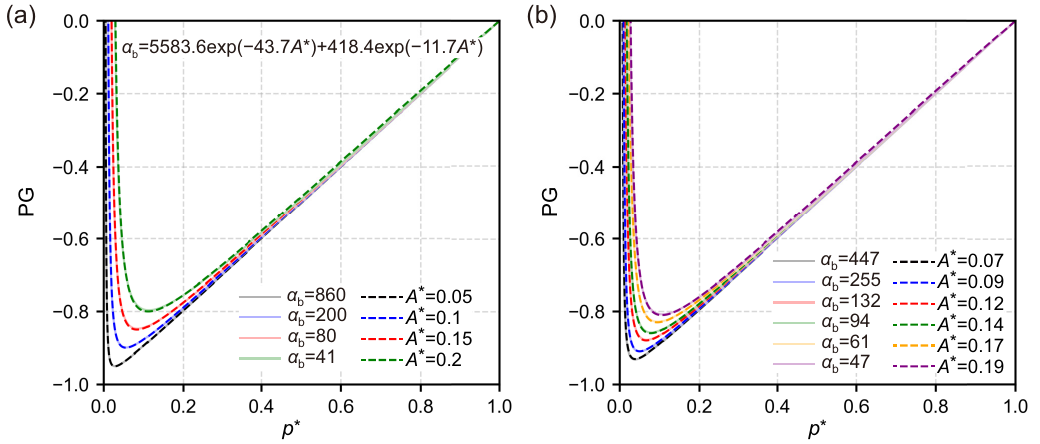}}
  \caption{Variation of PG with dimensionless outlet pressure $p^*$ for the check valve model and the sudden expansion model: (a) the backflow check strength $\alpha_b$ takes values of 41, 80, 200, and 860, corresponding to area ratios $A^* = 0.2, 0.15, 0.1, 0.05$, respectively; (b) by selecting several additional values of $A^*$, the corresponding $\alpha_b$ values are calculated based on the mapping relationship in (a).}
\label{fig5}
\end{figure}

Taking the sudden expansion model \citep{Fievisohn29} as an example, the feasibility of simulating the specific flow channel configuration with the check valve model is verified. When $\alpha_b > \alpha_1$, the gas in both the check valve model and the sudden expansion model exhibits the characteristic of flowing from the high-pressure region to the low-pressure region. Further analysis of the sudden expansion model reveals that the area ratio $A^*$ of its throat cross-section to the outlet cross-section can be used to characterize the degree of flow blocking. The ratio of the two models' outlet back pressure to the inlet total pressure is defined as the dimensionless pressure $p^*$. Figure~\ref{fig5}(a) presents the variation of PG with dimensionless outlet pressure $p^*$ for both models, under the conditions where $\alpha_b$ takes values of 41, 80, 200, and 860, respectively, and the corresponding $A^*$ values are 0.2, 0.15, 0.1, and 0.05. It can be seen that the PG calculation results of the two models are in good agreement. Therefore, a corresponding relationship can usually be established between $\alpha_b$ and the parameter used to define flow blocking for the specific flow channel configuration.

Further observation of the relationship between $\alpha_b$ and $A^*$ reveals that the former decreases monotonically with the latter. Additionally, the smaller the value of $A^*$, the faster the rate at which $\alpha_b$ changes with it. Based on the four known corresponding points of $\alpha_b$ and $A^*$, and considering their variation characteristics, the functional relationship between them is derived through algebraic operations as follows:

\begin{equation}
  \alpha_b=5583.6\exp\left(-43.7A^* \right)+418.4\exp\left(-11.7A^* \right)
  \label{a2.13}
\end{equation}

To verify the universality of the above mapping relationship, several additional values of $A^*$ are selected. After calculating the corresponding $\alpha_b$ values, the variation of the PG for both the check valve model and the sudden expansion model with the dimensionless outlet pressure $p^*$ is plotted in figure~\ref{fig5}(b). It can be seen that the results of the two models still maintain good consistency.

\section{Numerical method and physical model}
\label{sec3}

\subsection{Numerical method}
\label{sec3.1}

According to previous studies \citep{Schwer23,Yi31,Zhou32}, it is known that in the detonation flow field with high-speed flow, transport properties such as molecular viscosity, thermal conductivity, and mass diffusion have no significant impact on the flow field structure. Thus, the unsteady Euler equations incorporating chemical reaction source terms are adopted as the governing equations for numerical simulation in this study. Meanwhile, existing studies \citep{Sheng11,Yi31} have shown that accurately simulating the detonation wave system structure requires high grid resolution. Therefore, directly conducting 3D numerical simulation on the annular RDE with an expansion section will consume substantial computing resources. To address this issue, and considering the variable cross-section of the RDE, this study simplifies its radial features using the quasi-1D assumption along the axial direction and establishes the following governing equations for the quasi-2D RDE flow.

\begin{equation}
  \frac{\partial \rho}{\partial t} + \frac{\partial \rho u}{\partial x} + \frac{\partial \rho v}{\partial y} = -\rho v  \frac{1}{A}  \frac{\partial A}{\partial y}
  \label{a3.1}
\end{equation}
\begin{equation}
  \frac{\partial \rho u}{\partial t} + \frac{\partial (\rho u^2 + p)}{\partial x} + \frac{\partial \rho u v}{\partial y} = -\rho u v \frac{1}{A} \frac{\partial A}{\partial y}
  \label{a3.2}
\end{equation}
\begin{equation}
  \frac{\partial \rho v}{\partial t} + \frac{\partial \rho u v}{\partial x} + \frac{\partial (\rho v^2 + p)}{\partial y} = -\rho v^2 \frac{1}{A} \frac{\partial A}{\partial y}
  \label{a3.3}
\end{equation}
\begin{equation}
  \frac{\partial E}{\partial t} + \frac{\partial (E + p)u}{\partial x} + \frac{\partial (E + p)v}{\partial y} = -(E + p)v \frac{1}{A} \frac{\partial A}{\partial y}
  \label{a3.4}
\end{equation}
\begin{equation}
  \frac{\partial \rho Y_k}{\partial t} + \frac{\partial \rho u Y_k}{\partial x} + \frac{\partial \rho v Y_k}{\partial y} = \dot{w}_k - \rho v Y_k \frac{1}{A} \frac{\partial A}{\partial y}
  \label{a3.5}
\end{equation}
\begin{equation}
  p = \rho R T
  \label{a3.6}
\end{equation}

Where $Y_k$ is the mass fraction of the $k$-th species, $\dot{\omega}_k$ is the mass production rate of this species per unit volume, and $A$ is defined as the area ratio of the RDE's axial cross-section to its inlet cross-section. The expression for the total energy $E$ is given as follows:

\begin{equation}
  E = \sum_{k=1}^{N} \rho Y_k h_{a,k} - p + \frac{1}{2} \left( u^2 + v^2 \right)
  \label{a3.7}
\end{equation}

Where $h_{a,k}$ represents the absolute enthalpy of the $k$-th species.

To numerically solve the governing equations, this study adopts the JANC solver \citep{Wen33}, developed based on the JAX framework, and spatially discretizes equations (\ref{a3.1})$\sim$(\ref{a3.5}) using the finite difference method. For the discretization of convective terms, a scheme combining the KNP flux splitting format and the WENO5 flux reconstruction method is used. To address the coupling issue between convective terms and chemical reaction source terms in the governing equations, the Strang source term splitting method is employed. The time marching of convective terms uses the third-order TVD-Runge-Kutta scheme to ensure the accuracy and stability of temporal discretization, while the marching of chemical reaction source terms employs the first-order point implicit method, aiming to efficiently handle stiff chemical reaction kinetics. Numerical simulations are performed on the NVIDIA A100 GPU, with a time step set to $5 \times 10^{-9} \, \text{s}$ \citep{Wen33}. For the calculation of chemical reaction kinetics, a detailed hydrogen/air chemical reaction mechanism involving 9 species and 19 elementary reactions \citep{Choi34} is adopted.

\begin{figure}
  \centerline{\includegraphics[width=\textwidth]{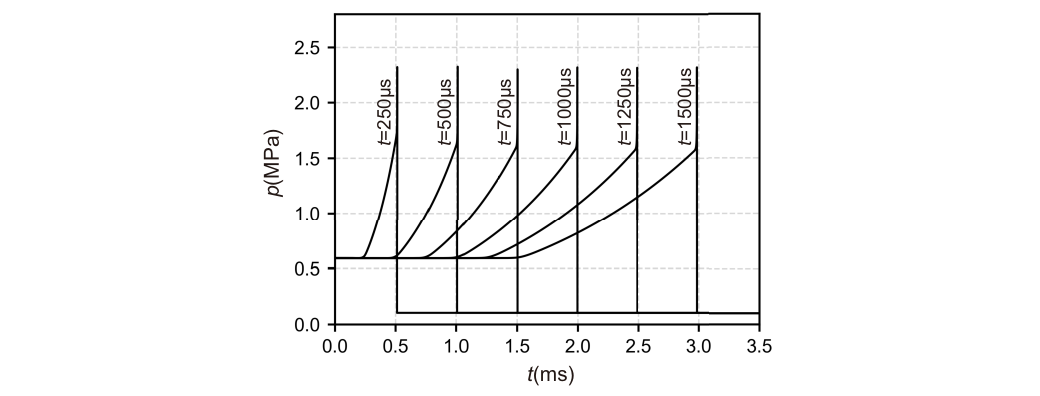}}
  \caption{Evolution of pressure distribution in the detonation tube at different times.}
\label{fig6}
\end{figure}

To verify the reliability of the numerical method, simulations are conducted on the flow field in a 3.5 m long 1D detonation tube with a grid size of 0.1 mm. The tube is filled with stoichiometric hydrogen/air premixed gas. The 5 mm range at the far left of the tube is set as the ignition zone, where the initial pressure and temperature are 5 MPa and 2000 K, respectively. The initial state of the remaining area outside this zone is $p_0 = 1$ atm and $T_0 = 300$ K. Figure~\ref{fig6} presents the evolutionary characteristics of the pressure distribution in the detonation tube at different times. The results show that this numerical method can correctly capture the 1D detonation wave structure. To further verify the method's accuracy quantitatively, table~\ref{tab1} presents a comparison between the numerical simulation results of detonation wave Chapman-Jouguet (C-J) parameters (velocity, pressure, temperature) and their theoretical values under different initial temperature conditions. Statistics show that the maximum error between numerical and theoretical results is less than 3\% for all cases. Therefore, this numerical method is capable of simulating the detonation phenomenon under the premixed hydrogen/air condition.

\begin{table}[h!] 
  \centering 
  \setlength{\tabcolsep}{6pt} 
  \caption{Numerical simulation results and theoretical values of detonation wave Chapman-Jouguet (C-J) parameters (velocity, pressure, temperature) under varying initial temperature conditions.}
  \label{tab1}
  \resizebox{\textwidth}{!}{ 
    \begin{tabular}{lcccccccc} 
      \toprule 
      $p_0$ (atm) & $T_0$ (K) & $U_{\text{CJ}}$ (m/s) & $U_{\text{num}}$ (m/s) & $P_{\text{CJ}}$ (atm) & $P_{\text{num}}$ (atm) & $T_{\text{CJ}}$ (K) & $T_{\text{num}}$ (K) & err (\%) \\[3pt]
      \midrule 
      1 & 300 & 1976.04 & 2000.05 & 15.54 & 15.41 & 2963.07 & 2959.68 & 1.22 \\
      1 & 600 & 1940.32 & 1950.05 & 7.77  & 7.55  & 3006.97 & 2996.65 & 2.83 \\
      1 & 900 & 1904.30 & 1925.05 & 5.18  & 5.05  & 3054.67 & 3046.53 & 2.51 \\
      \bottomrule 
    \end{tabular}
  }
\end{table}

\begin{figure}
  \centerline{\includegraphics[width=\textwidth]{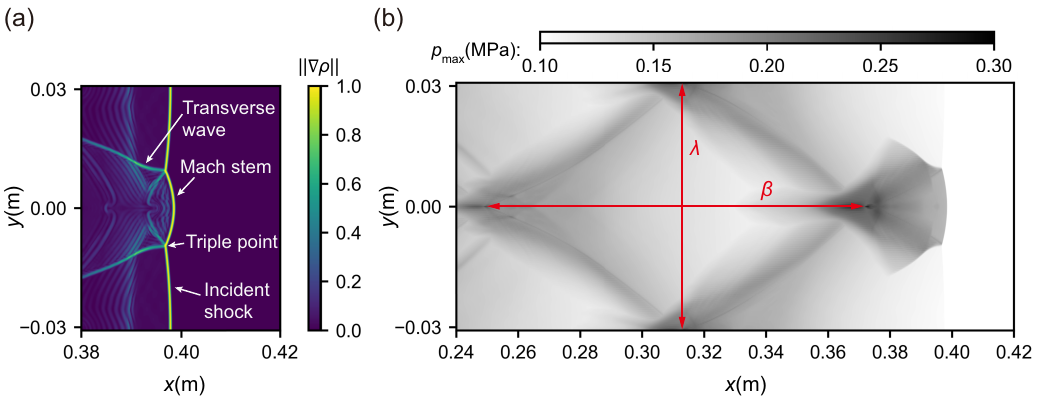}}
  \caption{(a) Normalized density gradient contour, $\|\nabla \rho\| = 1 - \exp\left( \frac{-5|\nabla \rho|}{|\nabla \rho|_{\text{max}}} \right)$; (b) Numerical schlieren image plotted through the pressure maximum.}
\label{fig7}
\end{figure}

To further verify the reliability of the numerical method, this study also conducts numerical simulations on a 2D detonation tube ($0.5\,\text{m} \times 0.0616\,\text{m}$) using the same initial conditions and grid resolution as those in references \citep{Marcantoni35}. The tube is filled with a premixed gas of hydrogen, oxygen, and argon, with a molar ratio of the species $\mathrm{H}_2 : \mathrm{O}_2 : \mathrm{Ar} = 2 : 1 : 7$, an initial pressure of $p_{\text{env}} = 6670\,\text{Pa}$, and an initial temperature of $T_{\text{env}} = 298\,\text{K}$. Three equally spaced high-temperature and high-pressure ignition zones are set on the left side of the tube. Each zone has a size of $3.072\,\text{mm} \times 8.8\,\text{mm}$, with an initial pressure of $p_I = 1000p_{\text{env}}$ and an initial temperature of $T_I = 25T_{\text{env}}$. The results are shown in figure~\ref{fig7}, where figure~\ref{fig7}(a) successfully captures the triple point structure of the detonation wave, including the incident shock wave, Mach stem, and transverse wave. Figure~\ref{fig7}(b) shows the numerical schlieren image plotted through the pressure maximum, in which the detonation cell aspect ratio $\lambda/\beta$ is 0.5, close to the result of 0.48 \citep{Marcantoni35}. The above analysis indicates that the proposed numerical method can accurately reproduce the key characteristics of the detonation wave, verifying its effectiveness.

\subsection{Physical model}
\label{sec3.2}

\begin{figure}
  \centerline{\includegraphics[width=\textwidth]{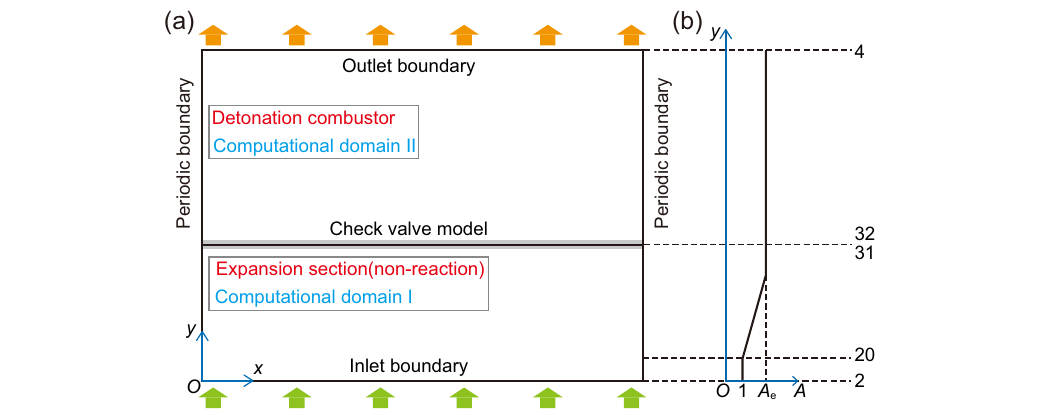}}
  \caption{(a) The quasi-2D RDE unfolded along the circumferential direction; (b) Distribution characteristics of the area ratio $A$ along the axial direction of the quasi-2D RDE.}
\label{fig8}
\end{figure}

The quasi-2D RDE unfolded along the circumferential direction is shown in figure~\ref{fig8}(a). It consists of three main parts: the variable cross-sectional expansion section, the detonation combustor, and the check valve model. Among these components, the combustor has a span length of 100 mm and an axial length of 38 mm, while the expansion section has an axial length of 22 mm. Figure~\ref{fig8}(b) shows the distribution features of the area ratio $A$ along the axial direction of the quasi-2D RDE. Station 2 is the inlet of the RDE, with a 1 mm long throat located downstream. Station 20 corresponds to the outlet of the throat, while station 31 corresponds to the exit of the expansion section. Upstream of station 31, there is a 2 mm long constant cross-section region. The area ratio $A_e$ at station 31 is defined as the expansion ratio of the RDE. Station 32 is the inlet of the combustor, and station 4 marks the outlet of the RDE.

The premixed stoichiometric hydrogen/air mixtures are injected into the expansion section through the inlet boundary. The total temperature and total pressure of the fresh gas at the inlet boundary are set to $T_{t,\text{in}} = 360\,\text{K}$ and $p_{t,\text{in}} = 1\,\text{MPa}$, respectively. To avoid shock wave reflection at the outlet boundary, a non-reflecting boundary condition is applied. Meanwhile, periodic boundaries are set on the left and right sides of the quasi-2D RDE to simulate the circumferential continuity of the annular combustor. The computational domain is divided into two independent regions by the check valve model, corresponding to computational domain I (expansion section) and domain II (detonation combustor) in figure~\ref{fig8}, respectively. These two regions are connected through the check valve model, whose upper and lower boundaries correspond to the combustor inlet and the expansion section outlet, respectively. Notably, the internal flow of the check valve model is not considered in this study.

It is assumed that no chemical reactions occur in the expansion section. When the fresh gas flows from the expansion section into the combustor (from station 31 to station 32), forward flow occurs in the check valve model; otherwise, backflow occurs. As concluded in section~\ref{sec2.2}, if the backflow check strength $\alpha_b$ is above $\alpha_1$, the flow can only proceed from the high-pressure region to the low-pressure region in order for the check valve model to have a physical solution. Therefore, backflow occurs when the combustor inlet pressure $p_{32}$ exceeds the expansion section outlet pressure $p_{31}$; otherwise, forward flow occurs. The specific solution process for the flow parameters at the upper and lower boundaries of the check valve model is as follows:

(1) When $p_{32} \geq p_{31}$, the gas flows from station 32 to station 31. During this process, the flow channel of the check valve model contracts, blocking the passing gas and leading to a discontinuous jump in the flow parameters at its upper and lower boundaries. First, parameters at the upper boundary of the check valve model are interpolated based on the flow parameters at station 32 (including pressure, temperature, spanwise velocity, species mass fractions, and specific heat ratio). Meanwhile, the pressure at the model's lower boundary is interpolated from the parameter at station 31. Then, the model's outlet-inlet pressure ratio can be determined as follows:

\begin{equation}
  y = \frac{p_{lb}}{p_{ub}} = \frac{p_{31}}{p_{32}}
  \label{a3.8}
\end{equation}

After appropriate transformation of Eq.~(2.9), a quadratic equation for the density ratio $x$ is obtained:

\begin{equation}
  ax^2 + bx + c = 0
  \label{a3.9}
\end{equation}

The coefficients of the equation are as follows:

\begin{equation}
  \begin{cases}
  a = \dfrac{\gamma+1}{2(\gamma-1)} \, y + \dfrac{1}{2} \\[6pt]
  b = \alpha_b \, \dfrac{\gamma}{\gamma-1} \, y - \dfrac{\gamma}{\gamma-1} \\[6pt]
  c = \dfrac{1}{2} \, y - \dfrac{1}{2} - \alpha_b \, \dfrac{\gamma}{\gamma-1}
  \end{cases}
  \label{a3.10}
\end{equation}

The root of the equation that satisfies $x \in (1, x_{g2})$ is as follows:

\begin{equation}
  x = \frac{-b + \sqrt{b^2 - 4ac}}{2a}
  \label{a3.11}
\end{equation}

Assuming that the mass fractions of each species at the upper and lower boundaries of the check valve model are the same, the temperature at the lower boundary of the model is then calculated using the ideal gas state equation:

\begin{equation}
  T_{lb} = xyT_{ub} = xyT_{32}
  \label{a3.12}
\end{equation}

To ensure that the check valve model strictly follows the conservation of momentum and energy, the spanwise velocities at its upper and lower boundaries must remain consistent. The calculation method for the axial velocities at these boundaries is as follows:

\begin{equation}
  v_{ub} = \sqrt{\frac{1}{\gamma} \frac{1-y}{\alpha_b + x}}* \sqrt{\frac{\gamma p_{ub}}{\rho_{ub}}} = \sqrt{\frac{1}{\gamma} \frac{1-y}{\alpha_b + x}} *\sqrt{\frac{\gamma p_{32}}{\rho_{32}}}, \quad v_{lb} = x v_{ub}
  \label{a3.13}
\end{equation}

(2) When $p_{32} < p_{31}$, the gas flows from station 31 to station 32. At this point, the flow channel of the check valve model is fully open, and under the isentropic assumption, all flow parameters at its upper and lower boundaries are completely consistent. Based on the axial velocity, static parameters, and specific heat ratio at station 31, the total pressure $p_{t,lb}$ and total density $\rho_{t,lb}$ at the lower boundary of the model can be calculated. According to this, the corresponding expression for the critical pressure $p_{cr}$ is as follows:

\begin{equation}
  p_{cr} = p_{t,lb} \left( \frac{2}{\gamma+1} \right)^{\frac{\gamma}{\gamma-1}}
  \label{a3.14}
\end{equation}

1) When $p_{cr} \leq p_{32} < p_{31}$, the fresh gas is injected into the combustor at subsonic speed.

\begin{equation}
  \begin{aligned}
  &p_{lb} = p_{ub} = p_{32}, \quad T_{lb} = T_{ub} = T_{31} \left( \frac{p_{ub}}{p_{31}} \right)^{\frac{\gamma-1}{\gamma}}, \\[6pt]
  &v_{lb} = v_{ub} = \sqrt{\frac{2\gamma}{\gamma-1}  \frac{p_{t,lb}}{\rho_{t,lb}} \left[ 1 - \left( \frac{p_{ub}}{p_{t,lb}} \right)^{\frac{\gamma-1}{\gamma}} \right]}
  \end{aligned}
  \label{a3.15}
\end{equation}

2) When $p_{32} < p_{cr}$, the fresh gas is injected into the combustor at sonic speed.

\begin{equation}
  \begin{aligned}
  &p_{lb} = p_{ub} = p_{cr}, \quad T_{lb} = T_{ub} = T_{31} \left( \frac{p_{ub}}{p_{31}} \right)^{\frac{\gamma-1}{\gamma}}, \\[6pt]
  &v_{lb} = v_{ub} = \sqrt{\frac{2\gamma}{\gamma-1}  \frac{p_{t,lb}}{\rho_{t,lb}} \left[ 1 - \left( \frac{p_{ub}}{p_{t,lb}} \right)^{\frac{\gamma-1}{\gamma}} \right]}
  \end{aligned}
  \label{a3.16}
\end{equation}

To achieve the coupled solution of computational domain I and II, it is first necessary to obtain the flow parameters at the upper and lower boundaries of the check valve model using the above method. Based on this, the values at the upper boundary are assigned to the virtual grids corresponding to the combustor inlet, while the values at the lower boundary are assigned to the virtual grids corresponding to the expansion section outlet. Through this boundary parameter assignment, the two-way transfer of flow field information between the two computational domains can be completed.

\subsection{Grid independence verification}
\label{sec3.3}

Table~\ref{tab2} systematically summarizes all the cases adopted in this study, covering multiple parameter combinations with different grid resolutions, expansion ratios $A_e$, and backflow check strengths $\alpha_b$. In particular, these cases include the extreme condition where $\alpha_b$ approaches infinity. It can be inferred from Eq. (2.10) that when $\alpha_b \to \infty$, the Mach number $M_{lb}$ at the lower boundary of the check valve model approaches 0. Therefore, backflow in the check valve model is completely blocked, and the model achieves the ideal check effect. In this case, the upper and lower boundaries of the model need to be treated as slip walls. Such a case can provide an important reference for exploring the extreme characteristics of the model.

\begin{table}[h!] 
  \centering 
  \caption{Summary of cases.}
  \label{tab2}
  \setlength{\tabcolsep}{8pt} 
  \begin{tabular}{lccc} 
    \toprule 
    Case & Check strength $\alpha_b$ & Expansion ratio $A_e$ & Grid size $\Delta x \times \Delta y$ (mm) \\[3pt]
    \midrule 
    1$\sim$5  & 20                       & 3.58, 4, 4.5, 5.5, 6.5 & 0.1$\times$0.1 \\
    6$\sim$11 & 41                       & 3.35, 3.58, 4, 4.5, 5.5, 6.5 & 0.1$\times$0.1 \\
    12$\sim$18& 200                      & 3, 3.35, 3.58, 4, 4.5, 5.5, 6.5 & 0.1$\times$0.1 \\
    19$\sim$26& $\infty$                 & 2.83, 3, 3.35, 3.58, 4, 4.5, 5.5, 6.5 & 0.1$\times$0.1 \\
    27        & 200                      & 4.5                     & 0.2$\times$0.2 \\
    28        & 200                      & 4.5                     & 0.05$\times$0.05 \\
    \bottomrule 
  \end{tabular}
\end{table}

\begin{figure}
  \centerline{\includegraphics[width=\textwidth]{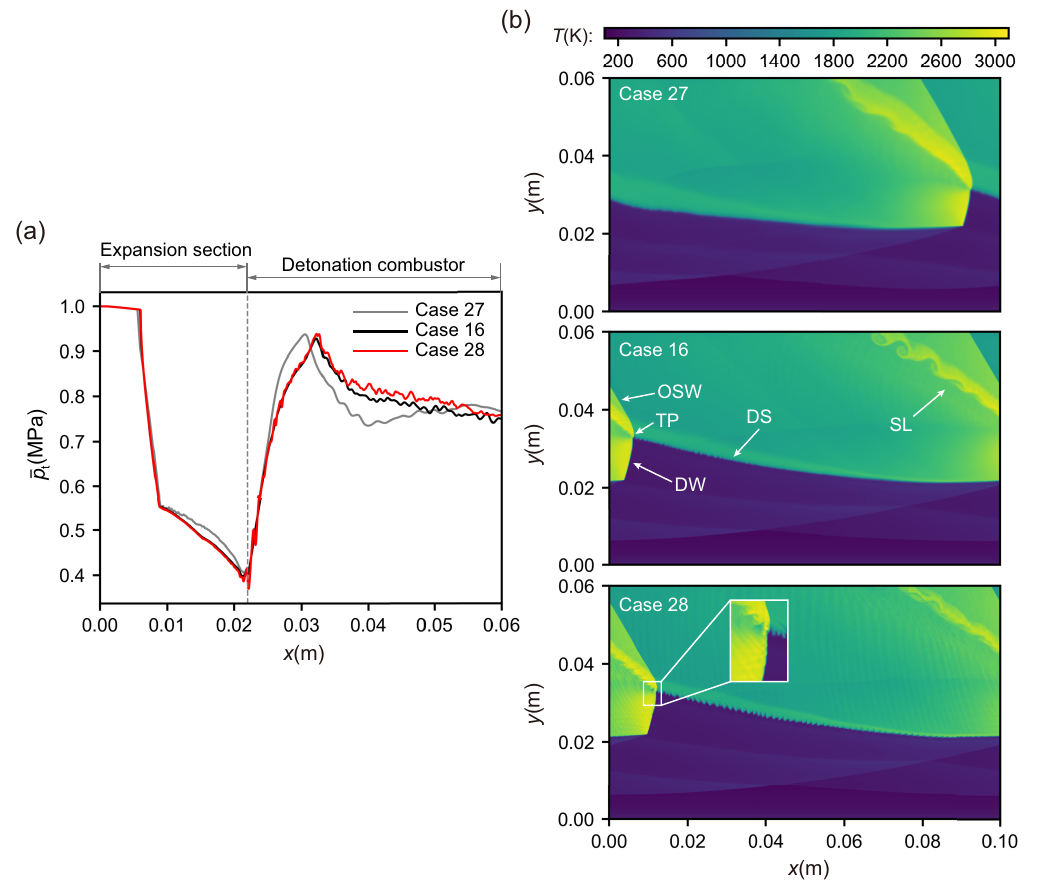}}
  \caption{(a) Axial distribution of mass-averaged total pressure $\bar{p}_t$ and (b) temperature contours of the internal flow field, both for the quasi-2D RDE under cases 16, 27, and 28 at $t = 2\,\text{ms}$.}
\label{fig9}
\end{figure}

This study selects case 16 ($A_e = 4.5$, $\alpha_b = 200$) as the baseline case. To investigate the influence of grid size on the numerical simulation results, comparative analyses with cases 16, 27, and 28 are shown in figure~\ref{fig9}. The mass-averaged total pressure $\bar{p}_t$ is defined as follows:

\begin{equation}
  \bar{p}_t = \frac{\int_0^L \rho v p_t \, dx}{\int_0^L \rho v \, dx}
  \label{a3.17}
\end{equation}

Where $L$ represents the spanwise length of the quasi-2D RDE, and $p_t$ denotes the total pressure calculated based on the axial velocity $v$. Figure~\ref{fig9}(a) presents the distribution characteristics of the mass-averaged total pressure $\bar{p}_t$ along the axial direction of the quasi-2D RDE under the three cases at $t = 2\,\text{ms}$. It can be seen that the distribution curves of $\bar{p}_t$ for case 16 and case 28 exhibit a high degree of consistency in their overall trends. The positions where the curves deviate in the expansion section are nearly identical, the points of minimum $\bar{p}_t$ in the expansion section almost overlap, and the points of maximum $\bar{p}_t$ in the combustor, as well as those at its outlet, also remain consistent. Only slight deviations exist in the axial descending phase at the end of the curves, with the overall oscillation amplitude for case 28 being slightly larger than that for case 16. In contrast, the $\bar{p}_t$ distribution for case 27 shows significant differences compared to the other two cases.

Figure~\ref{fig9}(b) shows the temperature contours of the quasi-2D RDE internal flow field for the three cases at $t=2\,\text{ms}$. From the results under different grid sizes, it can be observed that typical rotating detonation flow field structures—including the detonation wave (DW), deflagration surface (DS), slip line (SL), oblique shock wave (OSW), and triple point (TP)—are all present in the combustor of the quasi-2D RDE. Further analysis reveals that reducing the grid size significantly improves the simulation accuracy. High-resolution grids are able to capture flow field details more clearly and provide better resolution for detonation wave front instability, the Rayleigh-Taylor (R-T) instability on the deflagration surface, and the Kelvin-Helmholtz (K-H) instability of the slip line. This is likely the primary cause of the slight deviation in the $\bar{p}_t$ distribution between case 28 and case 16. Given that the core objective of this study is to investigate the effects of the expansion ratio $A_e$ and the backflow check strength $\alpha_b$ on the pressure-gain performance of the quasi-2D RDE, there is no need to overly pursue the flow field details enabled by high-resolution grids. Therefore, the grid size corresponding to case 16 ($\Delta x \times \Delta y = 0.1 \times 0.1\,\text{mm}$) is sufficient to meet the research requirements.

\section{Results and discussion}
\label{sec4}

\subsection{Typical flow structures}
\label{sec4.1}

\begin{figure}
  \centerline{\includegraphics[width=\textwidth]{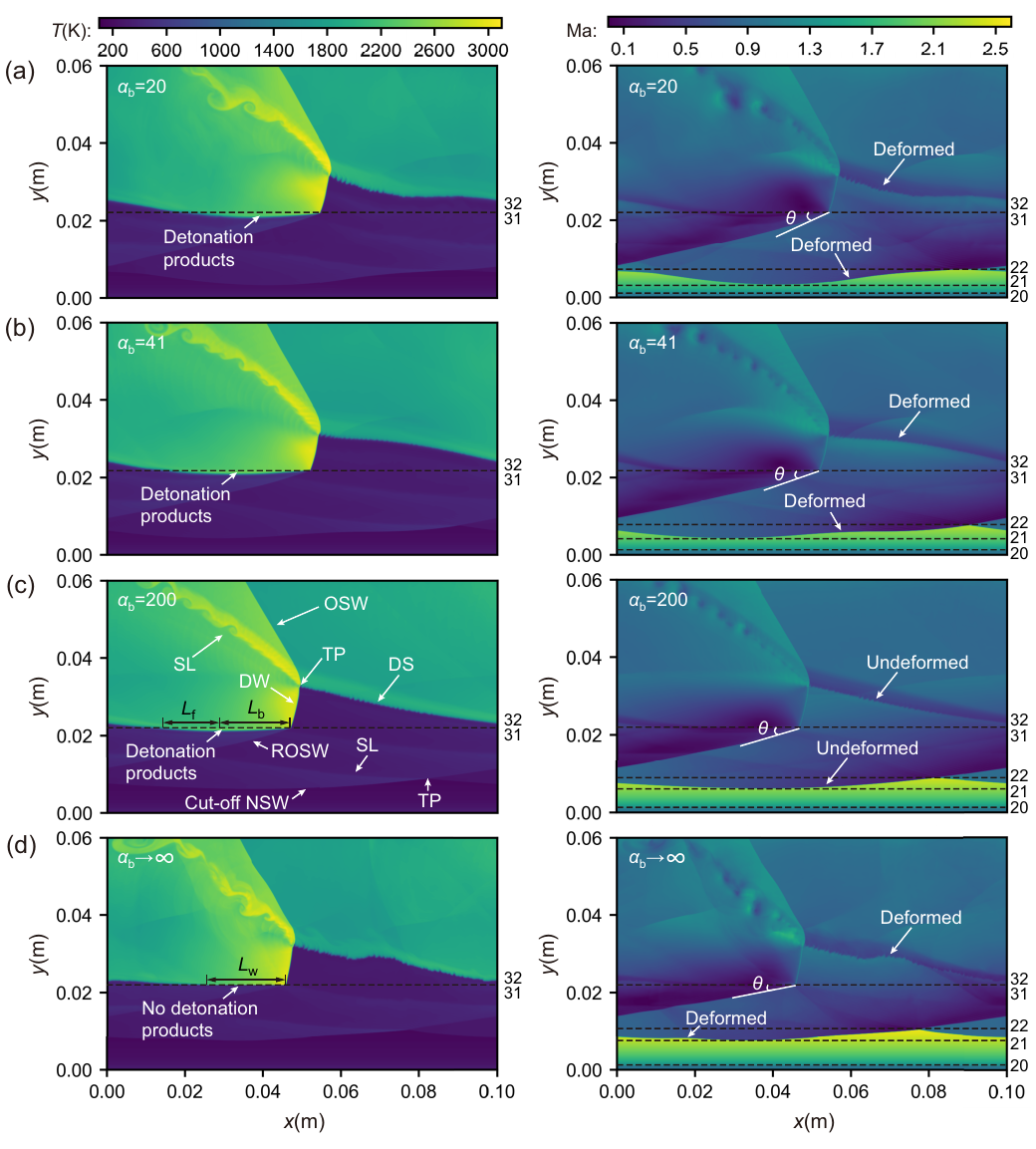}}
  \caption{Temperature and Mach number contours of the quasi-2D RDE internal flow field under different backflow check strengths $\alpha_b$ with the expansion ratio $A_e = 4.5$: (a) $\alpha_b = 20$, (b) $\alpha_b = 41$, (c) $\alpha_b = 200$, and (d) $\alpha_b \to \infty$.}
\label{fig10}
\end{figure}

Figure~\ref{fig10}(a) $\sim$ (d) present the temperature and Mach number contours of the quasi-2D RDE internal flow field corresponding to different backflow check strengths $\alpha_b$, under the condition of an expansion ratio $A_e = 4.5$. As observed from the temperature contour in figure~\ref{fig10}(c), in addition to the typical rotating detonation structure in the combustor, the flow field in the expansion section reveals that, influenced by the high-temperature and high-pressure region behind the detonation wave, a reverse oblique shock wave (ROSW) is generated at the lower boundary of the detonation wave, propagating upstream. Combined with the Mach number contour in figure~\ref{fig10}(c), it is evident that the flow at the throat reaches the speed of sound. After passing through station 20, the flow expands and accelerates to supersonic speeds as the area ratio $A$ increases. Driven by the supersonic incoming flow, the ROSW deflects its propagation direction at station 21, turning downstream in the expansion section. It eventually intersects with itself at station 22, forming a triple point and an accompanying slip line that propagates downstream.

Based on the above flow field characteristics, the back-propagating pressure between station 21 and station 22 is defined as the cut-off normal shock wave (cut-off NSW) in this study. This structure indicates that the back-propagating pressure is confined downstream of station 21 and cannot propagate to the engine inlet. Thus, it effectively prevents pressure disturbances from interfering with the intake process, which could otherwise lead to inlet blocking. This is a crucial premise for ensuring the validity of the calculation results for the quasi-2D RDE's total pressure recovery coefficient, as it maintains stability at the inlet boundary. Since the flow in the throat is sonic, it is essential to position station 21 above station 20; otherwise, the cut-off NSW that confines the back-propagating pressure cannot form.

There is a significant coupling correlation between the shape changes of the DS and the cut-off NSW in figure~\ref{fig10}. Specifically, when the DS exhibits obvious irregular fluctuations (as shown in figures~\ref{fig10}(a), (b), and (d)), the cut-off NSW also shows significant deformation. In contrast, the DS in figure~\ref{fig10}(c) remains a relatively smooth curved structure, and the corresponding cut-off NSW maintains a regular shape. This correlation may be related to the uneven energy release during detonation wave propagation.

As the backflow check strength $\alpha_b$ increases, the angle $\theta$ (marked in the Mach number contour of figure~\ref{fig10}) between the ROSW and the horizontal direction gradually decreases, causing station 21 to move downstream along the expansion section. This results in an increase in the Mach number ahead of the cut-off NSW and confines the back-propagating pressure to a region farther from the inlet. Due to the backflow of detonation products in the combustor, high-temperature regions appear below station 31 in the temperature contours. The sizes of these high-temperature regions vary under different $\alpha_b$ conditions. Notably, under the ideal check condition ($\alpha_b \to \infty$), no backflow occurs, but reverse shock wave structures still exist. This indicates that the check valve model is not a completely undisturbed control method. While it suppresses the back-propagating pressure, it inevitably perturbs the original flow field, and such perturbations cannot be entirely eliminated by increasing the backflow check strength.

\begin{figure}
  \centerline{\includegraphics[width=\textwidth]{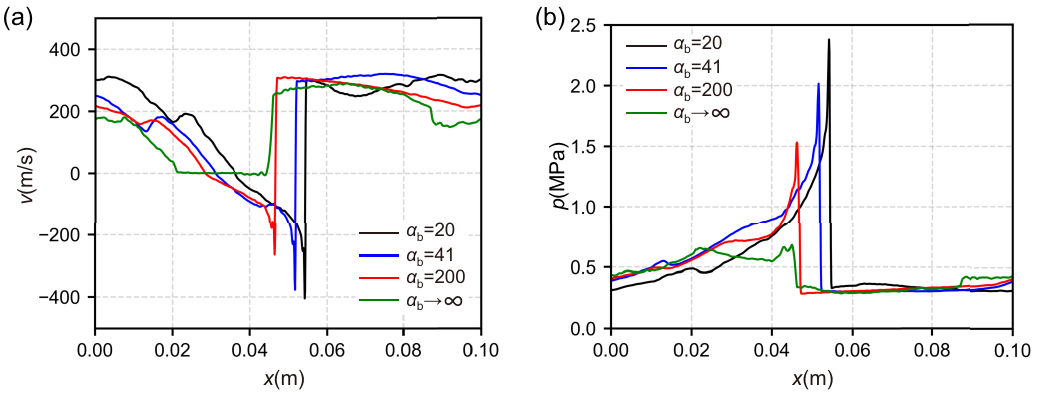}}
  \caption{Under the expansion ratio $A_e = 4.5$, the influence of different $\alpha_b$ values on the spanwise distribution of (a) pressure and (b) axial velocity at station 31.}
\label{fig11}
\end{figure}

Figures~\ref{fig11}(a) and (b) reveal the influence of different backflow check strengths $\alpha_b$ on the spanwise distribution of pressure and axial velocity at station 31, under the expansion ratio $A_e = 4.5$. As shown in figure~\ref{fig11}(a), as $\alpha_b$ increases, the absolute value of the backflow axial velocity peak at station 31 decreases. This change directly causes the angle $\theta$ to decrease gradually, and under the ideal check condition, $\theta$ reaches its minimum value. As shown in figure~\ref{fig11}(b), the pressure peak at station 31 declines with the increase of $\alpha_b$, reflecting the reduction in the intensity of the ROSW. Under the ideal check condition, the ROSW’s intensity reaches its minimum. This variation in intensity aligns with the change in the inclination angle $\theta$ and also confirms the positive effect of increasing the backflow check strength on suppressing back-propagating pressure.

\begin{figure}
  \centerline{\includegraphics[width=\textwidth]{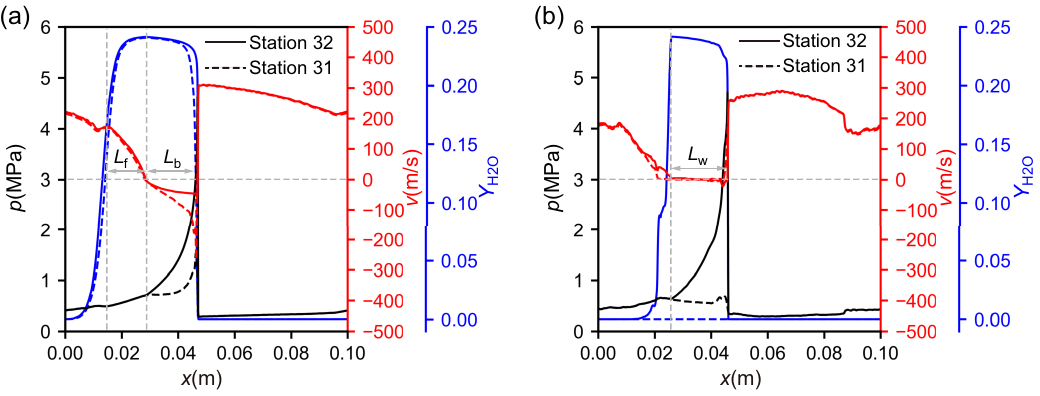}}
  \caption{Under the expansion ratio $A_e = 4.5$, (a) and (b) show the spanwise distribution of pressure, axial velocity, and water vapor mass fraction at stations 31 and 32, for backflow check strengths $\alpha_b = 200$ and $\alpha_b \to \infty$, respectively.}
\label{fig12}
\end{figure}

Under the expansion ratio $A_e = 4.5$, figures~\ref{fig12}(a) and (b) illustrate the spanwise distribution of pressure, axial velocity, and water vapor mass fraction at stations 31 and 32, for backflow check strengths $\alpha_b = 200$ and $\alpha_b \to \infty$, respectively. In figure~\ref{fig12}(a), region $L_b$ represents the area where detonation products undergo backflow. Within $L_b$, the pressure at station 32 is higher than at station 31, while the axial velocities at both stations are negative. Simultaneously, a relatively high water vapor mass fraction is observed at station 31, indicating that detonation products have flowed back into the expansion section via convection. Region $L_f$ represents the area where detonation products undergo forward flow. In $L_f$, the axial velocities at both stations 31 and 32 are positive; however, due to the earlier backflow of detonation products, the water vapor mass fraction in this region remains relatively high. As a result, fresh gas cannot be injected into the combustor. In figure~\ref{fig12}(b), region $L_w$ denotes the area of the slip wall under the ideal check condition. Within region $L_w$, the pressure at station 32 is higher than at station 31, and the axial velocities at both stations are zero. Fresh gas can be injected into the combustor in the other regions, as the axial velocity is positive and the water vapor mass fraction at station 31 remains zero. Regions $L_b$, $L_f$, and $L_w$ are also highlighted in the temperature contours shown in figures~\ref{fig10}(c) and (d). In summary, when the backflow check strength is $\alpha_b = 200$, the combustor’s intake block zone is $L_b + L_f$. For $\alpha_b \to \infty$, the block zone is $L_w$. Thus, under the non-ideal check condition, the injection of fresh gas is influenced by the pressure difference between the expansion section exit and the combustor inlet, as well as the backflow of detonation products. Consequently, the length of the intake block zone exceeds that of the backflow region. In contrast, under the ideal check condition, the injection of fresh gas depends only on the pressure difference.

\begin{figure}
  \centerline{\includegraphics[width=\textwidth]{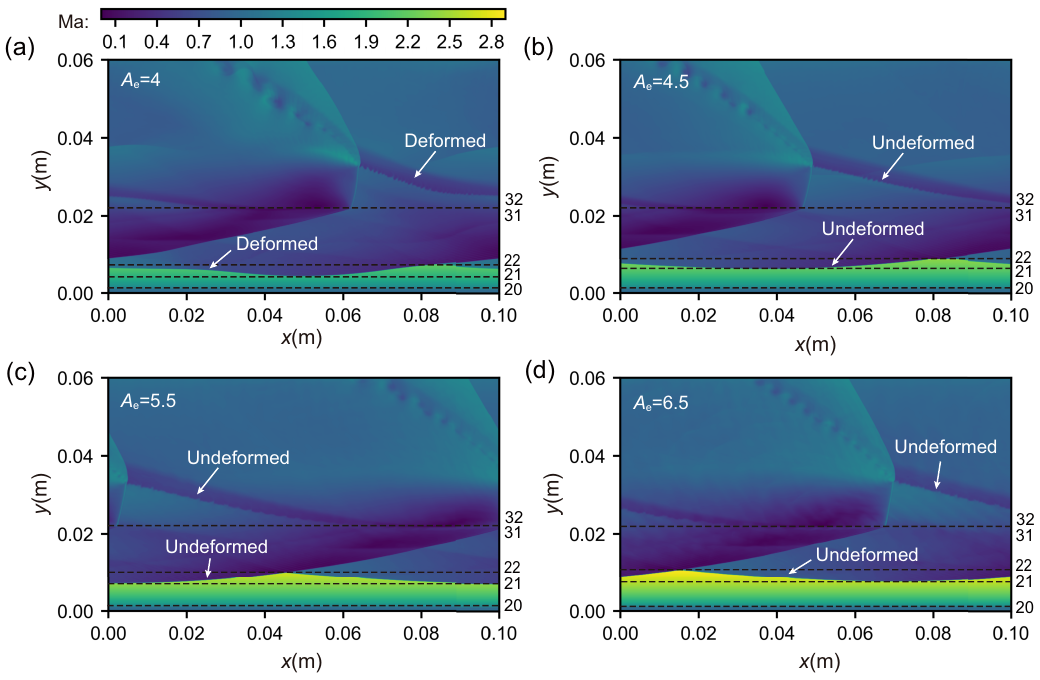}}
  \caption{Under the condition of backflow check strength $\alpha_b = 200$, Mach number contours of the quasi-2D RDE internal flow field for different expansion ratios $A_e$: (a) $A_e = 4$, (b) $A_e = 4.5$, (c) $A_e = 5.5$, and (d) $A_e = 6.5$.}
\label{fig13}
\end{figure}

Figure~\ref{fig13}(a) $\sim$ (d) present the Mach number contours of the quasi-2D RDE internal flow field corresponding to different expansion ratios $A_e$ under the condition of the backflow check strength $\alpha_b = 200$. It can be seen that as the expansion ratio $A_e$ increases, station 21 moves downstream in the expansion section. This occurs because the incoming flow can expand and accelerate more effectively as $A_e$ increases, which keeps the back-propagating pressure farther from the engine inlet. Meanwhile, under different expansion ratio conditions, there remains a clear correlation between the irregularity of the shapes of the DS and the cut-off NSW, reflecting the coupled interaction between both in the flow field of the quasi-2D RDE.

\subsection{Analysis of PG}
\label{sec4.2}

\begin{figure}
  \centerline{\includegraphics[width=\textwidth]{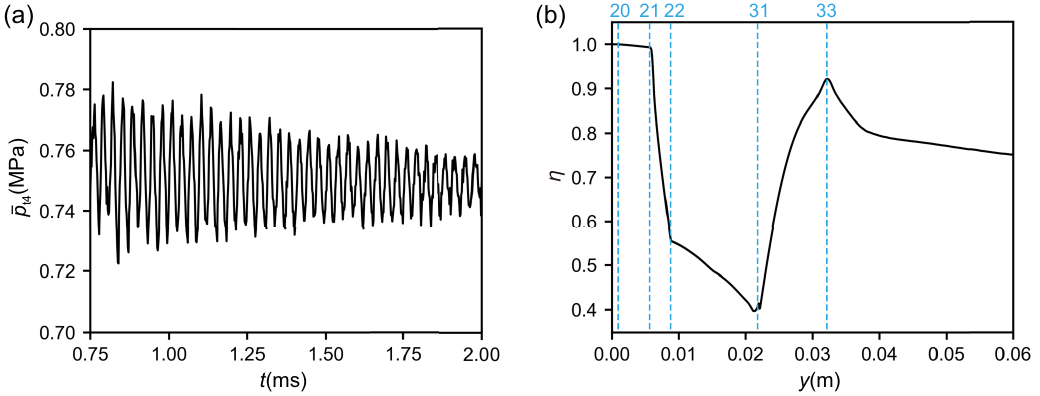}}
  \caption{For the quasi-2D RDE with expansion ratio $A_e = 4.5$ and backflow check strength $\alpha_b = 200$: (a) Time-varying characteristics of the mass-averaged total pressure $\bar{p}_{t4}$ at the outlet from $t_1 = 0.75\,\text{ms}$ to $t_2 = 2\,\text{ms}$; (b) Axial distribution of the total pressure recovery coefficient $\eta$ at different stations.}
\label{fig14}
\end{figure}

When the expansion ratio $A_e = 4.5$ and the backflow check strength $\alpha_b = 200$, the time-varying characteristics of the mass-averaged total pressure $\bar{p}_{t4}$ at the quasi-2D RDE outlet for the interval from $t_1 = 0.75\,\text{ms}$ to $t_2 = 2\,\text{ms}$ are shown in figure~\ref{fig14}(a). It can be seen that $\bar{p}_{t4}$ exhibits oscillations as the detonation wave propagates in the combustor. Given the strong transient behavior of $\bar{p}_{t4}$, this study adopts the time-averaging method to process $\bar{p}_t$ within the period from $t_1$ to $t_2$. Based on this, the axial distribution of the total pressure recovery coefficient $\eta$ at different stations of the RDE is shown in figure~\ref{fig14}(b), where $\eta$ is defined as follows:

\begin{equation}
  \eta = \frac{1}{p_{t,\text{in}}} \frac{1}{t_2 - t_1} \int_{t_1}^{t_2} \bar{p}_t \, dt
  \label{a4.1}
\end{equation}

The axial positions marked by the blue dashed lines in figure~\ref{fig14}(b) correspond to the stations labeled in the Mach number contour of figure~\ref{fig10}(c). From this, we can see that in the throat upstream of station 20, $\eta$ remains constant at 1, indicating no total pressure loss in this region. From station 20 to station 21, $\eta$ shows a slight downward trend. After entering station 21, affected by the cut-off NSW, $\eta$ begins to decrease rapidly until station 22. Starting from station 22, $\eta$ continues to decrease under the influence of the ROSW, but its rate of decrease slows down significantly compared to the previous stage. When the flow reaches station 31, fresh gas is injected into the combustor through the check valve model. Driven by the intense energy release from detonation combustion, the gas total pressure increases, and $\eta$ rises accordingly, reaching its maximum value at station 33—the location of the TP in the combustor. After that, $\eta$ gradually decreases again under the action of the OSW, with a value of 0.751 at the RDE exit.

\begin{figure}
  \centerline{\includegraphics[width=\textwidth]{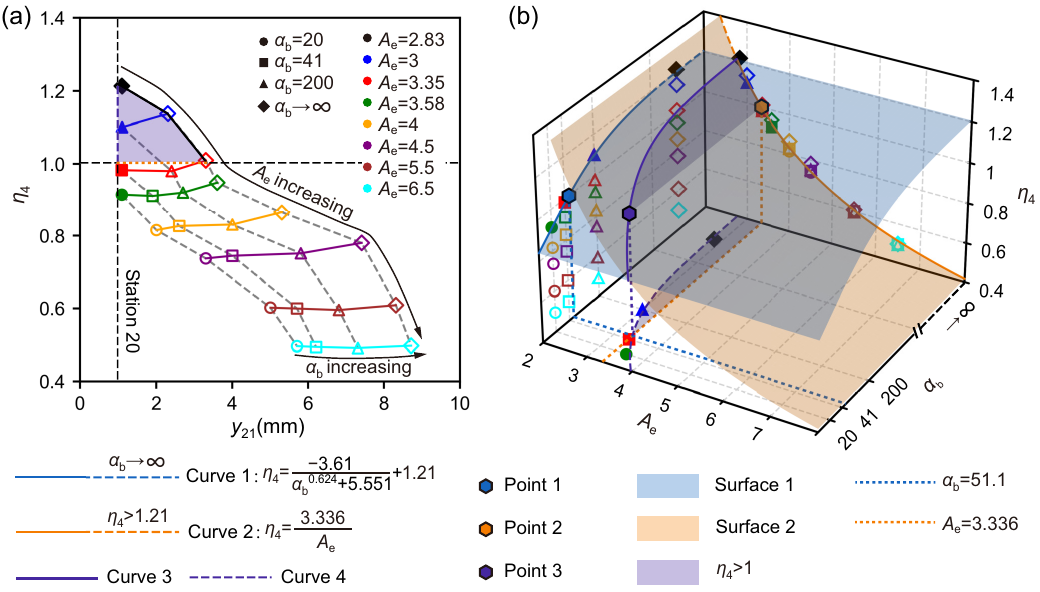}}
  \caption{(a) Distribution of the total pressure recovery coefficient $\eta_4$ at the exit of the RDE with the axial position $y_{21}$ of station 21 under different expansion ratios $A_e$ and backflow check strengths $\alpha_b$; (b) Three coordinate planes show the variation of $\eta_4$ with $\alpha_b$, the variation of $\eta_4$ with $A_e$, and the corresponding relationship between the minimum $A_e$ and $\alpha_b$. The solid points correspond to the cases where the back-propagating pressure just reaches the throat exit at station 20; curve 1 is the fitting curve of the solid points in the $\alpha_b$-$\eta_4$ plane, with its dashed portion representing the asymptote $\eta_4 = 1.21$ as $\alpha_b \to \infty$; curve 2 is the fitting curve of data points in the $A_e$-$\eta_4$ plane, with its dashed portion representing $\eta_4$ exceeding 1.21; surface 1 and surface 2 are the stretched surfaces of curve 1 along the $A_e$-axis and curve 2 along the $\alpha_b$-axis, respectively; curve 3 is the intersection of surface 1 and surface 2; points 1, 2, and 3 are the intersections of the plane $\eta_4 = 1$ with curves 1, 2, and 3, respectively; curve 4 is the projection of curve 3 onto the $\alpha_b$-$A_e$ plane; the purple region represents the area where $\eta_4 > 1$.}
\label{fig15}
\end{figure}

The axial position of station 21 in figure~\ref{fig14}(b) is defined as $y_{21}$, and the total pressure recovery coefficient at the RDE exit is defined as $\eta_4$. Under different expansion ratios $A_e$ and backflow check strengths $\alpha_b$, figure~\ref{fig15}(a) shows the distribution of the total pressure recovery coefficient $\eta_4$ with respect to the axial position $y_{21}$. From this, we can see that when $\alpha_b$ remains constant, as $A_e$ increases, $y_{21}$ increases accordingly. Meanwhile, $\eta_4$ decreases significantly, implying an increase in the total pressure loss at the RDE exit. We can also observe that when $A_e > 4$, the rate at which $y_{21}$ increases with $A_e$ gradually slows down. It is inferred that once $A_e$ reaches a specific value, $y_{21}$ may cease to increase further. However, a larger $A_e$ will further reduce $\eta_4$ by enhancing the intensity of the cut-off NSW, as will be explained subsequently. Therefore, since $\eta_4$ is already quite small when $A_e$ is relatively large, there is no need to conduct further research on the specific $A_e$ value at which $y_{21}$ tends to stabilize. When $A_e$ is kept constant, $y_{21}$ also rises as $\alpha_b$ increases; however, $\eta_4$ does not exhibit significant changes. Additionally, as $A_e$ increases, the range over which $\alpha_b$ controls $y_{21}$ first expands and then contracts.

In figure~\ref{fig15}(a), the solid dots correspond to cases where the back-propagating pressure just reaches the throat exit at station 20. As described in section~\ref{sec4.1}, this represents the most upstream location attainable by the back-propagating pressure. The observed variation law shows that, for a given backflow check strength $\alpha_b$, $\eta_4$ decreases monotonically with $y_{21}$, while $A_e$ increases monotonically with $y_{21}$. This indicates that as $y_{21}$ approaches station 20, the quasi-2D RDE achieves the maximum total pressure recovery coefficient and the minimum expansion ratio for the corresponding $\alpha_b$. At this point, any further reduction in $A_e$ will result in inlet blocking. For a fixed $y_{21}$, $\eta_4$ increases with $\alpha_b$ and reaches its maximum under the ideal check condition. Consequently, the quasi-2D RDE attains a global maximum total pressure recovery coefficient of 1.21 when $A_e = 2.83$ and $\alpha_b \to \infty$.

In figure~\ref{fig15}(b), the three coordinate planes show the variation of $\eta_4$ with $\alpha_b$, the variation of $\eta_4$ with $A_e$, and the corresponding relationship between the minimum value of $A_e$ and $\alpha_b$. In the $\alpha_b$-$\eta_4$ plane, as $\alpha_b$ increases, the maximum achievable $\eta_4$ of the RDE increases, but the rate of growth gradually slows down. The expression for curve 1, obtained by fitting the solid dot data, is given in equation (\ref{a4.2}). The dashed portion of curve 1 represents the asymptote $\eta_4 = 1.21$ as $\alpha_b \to \infty$, which corresponds to the global maximum total pressure recovery coefficient obtained in this study.

\begin{equation}
  \eta_4 = \frac{-3.61}{\alpha_b^{0.624} + 5.551} + 1.21
  \label{a4.2}
\end{equation}

In the $A_e$-$\eta_4$ plane, regardless of $\alpha_b$, $\eta_4$ increases as $A_e$ decreases; moreover, smaller values of $A_e$ correspond to a faster growth rate of $\eta_4$. This variation characteristic is consistent with the numerical simulation results by \citet{Kaemming10}. The expression for curve 2, obtained by fitting the data points in the plane, is given in equation (\ref{a4.3}). The mean square error of the fitting curve is $\sigma^2 = 3.97 \times 10^{-4}$. The $\eta_4$ values represented by the dashed portion of curve 2 exceed the maximum value of 1.21. It is possible that, based on this study, optimizing the geometric configuration of the quasi-2D RDE could enable it to achieve $\eta_4$ at smaller values of $A_e$, thereby realizing the dashed portion of curve 2.

\begin{equation}
  \eta_4 = \frac{3.336}{A_e}
  \label{a4.3}
\end{equation}

By stretching curve 1 along the $A_e$-axis and curve 2 along the $\alpha_b$-axis, surface 1 and surface 2 are obtained, respectively. The intersection of these two surfaces yields curve 3. The intersections of the plane $\eta_4 = 1$ with curves 1, 2, and 3 are denoted as points 1, 2, and 3, respectively. From point 1, the critical value of $\alpha_b$, denoted as $\alpha_{b,\text{cr}} = 51.1$, can be derived. When $\alpha_b > \alpha_{b,\text{cr}}$, the maximum PG of the quasi-2D RDE is positive. In contrast, if $\alpha_b < \alpha_{b,\text{cr}}$, the PG remains negative. From point 2, the critical value of $A_e$ is found to be $A_{e,\text{cr}} = 3.336$. When $A_e < A_{e,\text{cr}}$, the PG is guaranteed to be positive. However, if $A_e$ cannot be reduced below this critical value, simply adjusting the backflow check strength will not allow the RDE to attain a positive PG. Curve 4 is the projection of curve 3 into the $\alpha_b$-$A_e$ plane. This curve depicts the restrictive relationship between $\alpha_b$ and the minimum $A_e$ required to achieve positive PG. Specifically, only when $\alpha_b > \alpha_{b,\text{cr}}$ can the minimum $A_e$ be reduced below the critical value $A_{e,\text{cr}}$ without causing inlet blocking. The purple regions in figures~\ref{fig15}(a) and (b) represent the areas where $\eta_4 > 1$. It can be inferred from this that as $\alpha_b$ increases, the quasi-2D RDE can achieve positive PG over a wider range of $A_e$.

The above results indicate that effectively suppressing the back-propagating pressure is crucial for positive PG. The absence of such suppression may be the primary reason for the negative PG observed in RDE experiments, such as those by \citet{Bach13}. Therefore, under current experimental conditions, enhancing this suppression by optimizing the flow channel configuration represents a core aerodynamic challenge for enabling RDEs to achieve positive PG. This finding identifies the direction for adjusting key parameters to enhance RDE pressure-gain performance in the future.

\subsection{Total pressure recovery in the expansion section}
\label{sec4.3}

Figure~\ref{fig14}(b) illustrates that within the expansion section, the total pressure recovery coefficient $\eta$ decreases in two distinct stages: from station 21 to station 22, and from station 22 to station 23. The total pressure recovery coefficient for the first stage, induced by the cut-off NSW, is denoted as $\eta_a$. The total pressure recovery coefficient for the second stage, induced by the ROSW, is denoted as $\eta_b$. The methods for calculating $\eta_a$ and $\eta_b$ are presented as follows:

\begin{equation*}
  \eta_a = \frac{\eta_{22}}{\eta_{21}},\eta_b = \frac{\eta_{31}}{\eta_{22}}
  \label{Helm}
\end{equation*}

Where, $\eta_{21}$, $\eta_{22}$, and $\eta_{23}$ represent the total pressure recovery coefficients at stations 21, 22, and 23, respectively.

\begin{figure}
  \centerline{\includegraphics[width=\textwidth]{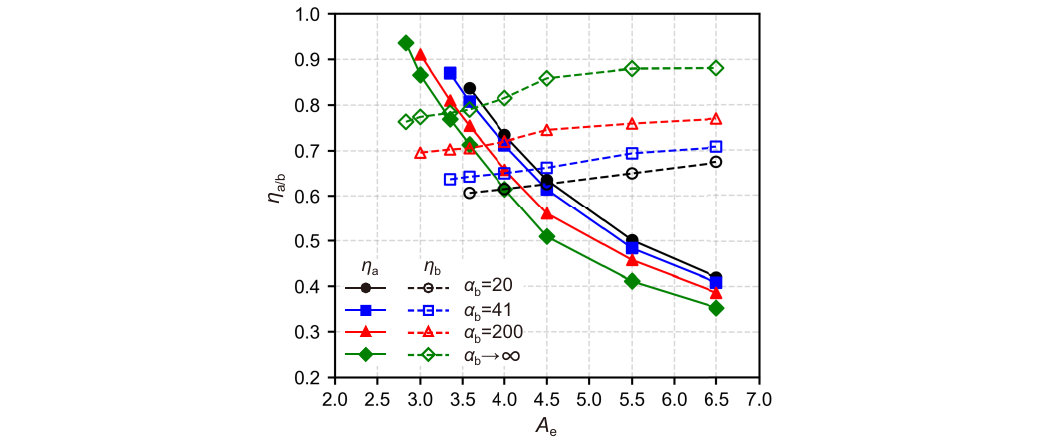}}
  \caption{The evolution of total pressure recovery coefficients $\eta_a$ (induced by the cut-off NSW) and $\eta_b$ (induced by the ROSW) with the expansion ratio $A_e$ under different backflow check strengths $\alpha_b$. The solid lines correspond to the variation of $\eta_a$ with $A_e$, and the dashed lines correspond to the variation of $\eta_b$ with $A_e$.}
\label{fig16}
\end{figure}

Figure~\ref{fig16} shows the evolution of $\eta_a$ and $\eta_b$ with the expansion ratio $A_e$ under different backflow check strengths $\alpha_b$. The solid lines correspond to the variation of $\eta_a$ with $A_e$, while the dashed lines represent the variation of $\eta_b$ with $A_e$. Specifically, with $\alpha_b$ fixed, $\eta_a$ decreases significantly as the expansion ratio $A_e$ increases, while $\eta_b$ rises slowly. When $A_e$ is held constant, an increase in $\alpha_b$ results in a reduction in $\eta_a$ and an increase in $\eta_b$, with the numerical fluctuation range of $\eta_b$ being notably larger than that of $\eta_a$. These results indicate that the cut-off NSW total pressure recovery coefficient $\eta_a$ is primarily influenced by the expansion ratio $A_e$, while the ROSW total pressure recovery coefficient $\eta_b$ is mainly affected by the backflow check strength $\alpha_b$.

\begin{figure}
  \centerline{\includegraphics[width=\textwidth]{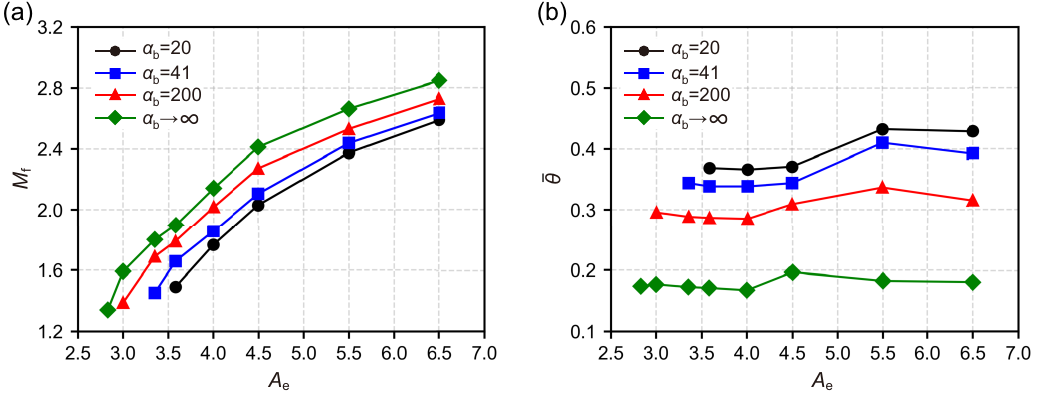}}
  \caption{Under different backflow check strengths $\alpha_b$: (a) Variation of the Mach number $M_f$ ahead of the cut-off NSW with the expansion ratio $A_e$; (b) Variation of the time-averaged angle $\bar{\theta}$ (between the ROSW and the horizontal direction) over the time range from 0.75 ms to 2 ms with the expansion ratio $A_e$.}
\label{fig17}
\end{figure}

To further explore the mechanisms by which the expansion ratio $A_e$ and backflow check strength $\alpha_b$ affect the total pressure recovery coefficients $\eta_a$ and $\eta_b$, figure~\ref{fig17} systematically presents the development of key flow field parameters in the quasi-2D RDE with respect to $A_e$ under different $\alpha_b$ conditions. Specifically, figure~\ref{fig17}(a) shows the variation of the Mach number $M_f$ ahead of the cut-off NSW, while figure~\ref{fig17}(b) illustrates the variation of the time-averaged angle $\bar{\theta}$ (between the ROSW and the horizontal direction) over the time range from 0.75 ms to 2 ms. Based on the definition of the cut-off NSW provided in section~\ref{sec4.1}, the flow parameters ahead of this shock wave are non-uniform. Therefore, this study defines the Mach number $M_f$ ahead of the cut-off NSW as follows: assuming the gas fully expands in the expansion section, the theoretical Mach numbers $M_{21}$ and $M_{22}$ are calculated based on the locations of stations 21 and 22, obtained from the axial distribution of $\eta$ shown in figure~\ref{fig14}(b). The arithmetic mean of these two values, $(M_{21}+M_{22})/2$, is then used as the approximate value for $M_f$. As seen in figure~\ref{fig17}(a), when $\alpha_b$ is fixed, $M_f$ increases with $A_e$, which intensifies the cut-off NSW and results in a decrease in the total pressure recovery coefficient $\eta_a$. However, when $A_e$ is fixed, changes in $\alpha_b$ cause $M_f$ to fluctuate only within a narrow range, indicating that the backflow check strength has a much weaker influence on $M_f$ compared to the expansion ratio. In figure~\ref{fig17}(b), when $\alpha_b$ is held constant, the time-averaged angle $\bar{\theta}$ is less affected by changes in $A_e$. However, when $A_e$ is fixed, variations in $\alpha_b$ cause $\bar{\theta}$ to change over a relatively wide range, with $\bar{\theta}$ decreasing as $\alpha_b$ increases. The reduction in the inclination angle weakens the intensity of the ROSW, which in turn increases the total pressure recovery coefficient $\eta_b$.

\begin{figure}
  \centerline{\includegraphics[width=\textwidth]{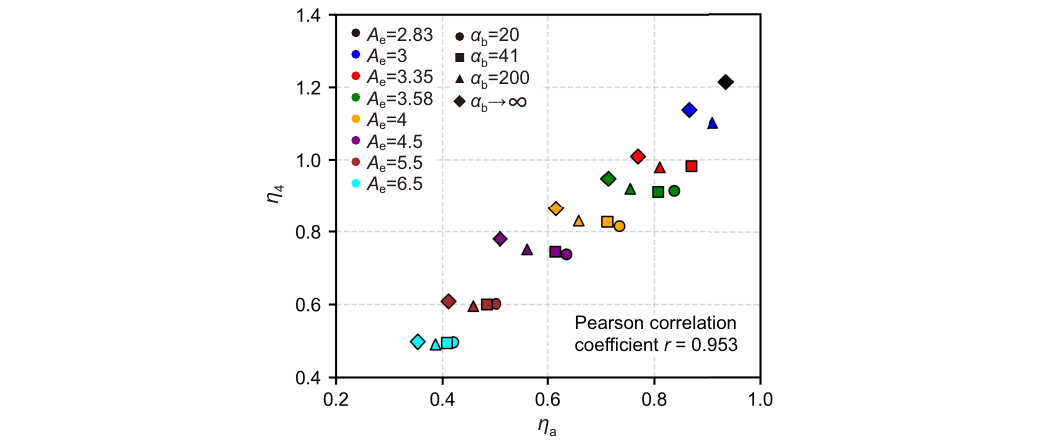}}
  \caption{Correlation between the total pressure recovery coefficient $\eta_4$ (at the RDE exit) and $\eta_a$ (induced by the cut-off NSW) under different expansion ratios $A_e$ and backflow check strengths $\alpha_b$.}
\label{fig18}
\end{figure}

From the analysis of the effects of expansion ratios and backflow check strengths on the total pressure recovery coefficients $\eta_4$, $\eta_a$, and $\eta_b$, it can be concluded that both $\eta_4$ and $\eta_a$ are primarily influenced by $A_e$. Therefore, the dominant contributor to the total pressure loss at the exit of the quasi-2D RDE is the loss induced by the cut-off NSW. Figure~\ref{fig18} illustrates the relationship between $\eta_4$ and $\eta_a$ under various expansion ratios $A_e$ and backflow check strengths $\alpha_b$. The Pearson linear correlation coefficient between $\eta_a$ and $\eta_4$ is calculated to be $r = 0.953$, indicating a strong positive linear correlation between the two. This result further supports, from a statistical perspective, that the total pressure recovery due to the cut-off NSW is the critical factor influencing the pressure-gain performance of the RDE.

In summary, when the expansion ratio is small, the intensity of the cut-off NSW is weak. At this time, enhancing the backflow check strength to prevent inlet blocking is crucial for maintaining positive PG. Therefore, optimizing the flow channel to efficiently suppress back-propagating pressure is a beneficial approach to improving the RDE’s PG.

\subsection{General PG criterion for RDEs}
\label{sec4.4}

\begin{figure}
  \centerline{\includegraphics[width=\textwidth]{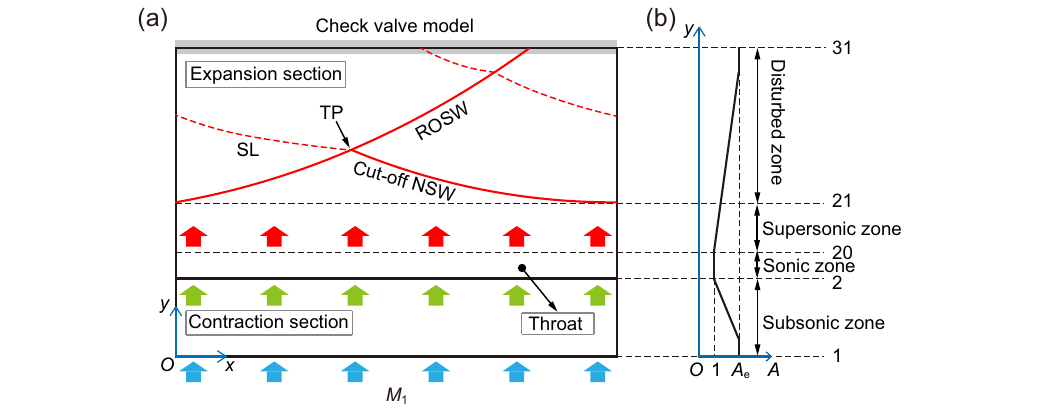}}
  \caption{(a) The circumferentially unfolded structure of the normalized physical model and (b) the distribution of the area ratio $A$ along its axial direction.}
\label{fig19}
\end{figure}

To eliminate the impact of differences in the expansion ratio $A_e$ on the pressure-gain performance, normalization of the quasi-2D RDE is required. The specific process is as follows: based on the original expansion section, an additional contraction section is added upstream of the throat, with an inlet cross-sectional area equal to the outlet of the RDE. The area ratio of the contraction section's inlet cross-section to the throat's cross-section is defined as $A_1$. Since the combustor of the quasi-2D RDE has a constant cross-section, the condition $A_e/A_1 = 1$ is satisfied. The circumferentially unfolded structure of the normalized physical model and the distribution of the area ratio $A$ along its axial direction are shown in figure~\ref{fig19}(a) and (b), respectively, where station 1 corresponds to the inlet of the additional contraction section. According to section~\ref{sec4.1}, the flow in the throat is sonic, transitioning to supersonic in the downstream expansion section. Further inference based on gas dynamics theory reveals that the flow in the contraction section is subsonic. The sonic surface at the throat plays a critical role in isolating back-propagating pressure downstream of station 21. If such pressure disturbances propagate through the sonic surface, they will directly interfere with the upstream components. Therefore, the throat structure is essential for ensuring the continuous operation of the RDE's working cycle.

\begin{figure}
  \centerline{\includegraphics[width=\textwidth]{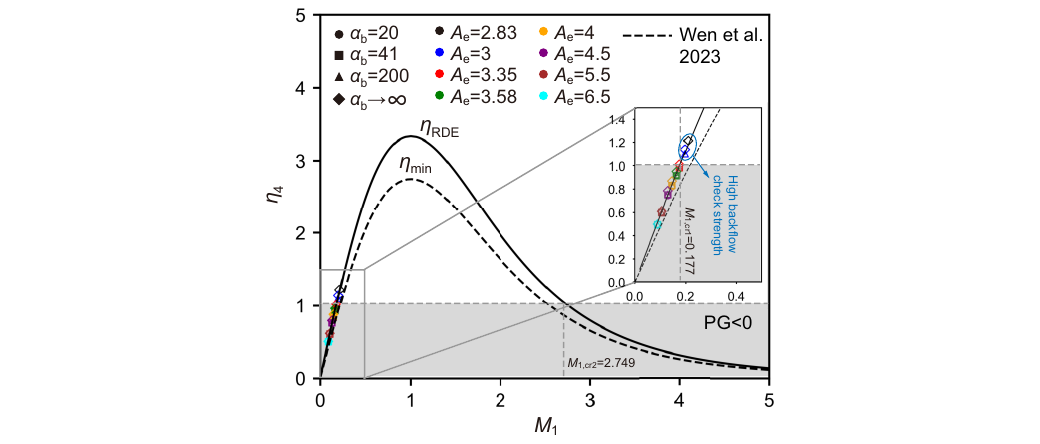}}
  \caption{When the reactants are stoichiometric hydrogen/air mixtures: the scatter points show the variation of the total pressure recovery coefficient $\eta_4$ at the exit of the quasi-2D RDE with the equivalent inlet Mach number $M_1$; $\eta_{\text{RDE}}$ is the curve of $\eta_4$ varying with $M_1$ obtained through the normalization of the quasi-2D RDE based on the functional relationship $\eta_4 = 3.336 / A_e$ derived in section~\ref{sec4.2}; $\eta_{\text{min}}$ is the curve of $\eta_4$ varying with $M_1$ obtained through theoretical derivation according to references \citep{Wen22}.}
\label{fig20}
\end{figure}

Under the isentropic assumption, the equivalent inlet Mach number $M_1$ at station 1 can be derived from the area ratio $A_1$. This parameter, after normalization, can serve as a core characteristic parameter for subsequent analyses, replacing the specific expansion ratio $A_e$ in the original physical model. The scatter points in fgure~\ref{fig20} illustrate the variation of the total pressure recovery coefficient $\eta_4$ at the exit of the quasi-2D RDE with the equivalent inlet Mach number $M_1$ for all cases in this study. It can be seen that $M_1$ is less than 1, and regardless of the backflow check strength, $\eta_4$ increases with $M_1$. Based on the functional relationship $\eta_4 = 3.336/A_e$ derived in section~\ref{sec4.2}, the variation curve of $\eta_4$ with $M_1$ (denoted as $\eta_{RDE}$) can be obtained through the normalization of the quasi-2D RDE, as shown in figure~\ref{fig20}. The critical Mach numbers corresponding to $\eta_{RDE} = 1$ are $M_{1,cr1} = 0.177$ and $M_{1,cr2} = 2.749$, respectively. Since all scatter points to the right of $M_{1,cr1}$ correspond to relatively high $\alpha_b$, a larger backflow check strength is key for the quasi-2D RDE to achieve positive PG when the inlet Mach number $M_1 > M_{1,cr1}$. Otherwise, the upstream components will be affected by back-propagating pressure.

Based on the assumptions in references \citep{Wen22} that the exist flow of the pressure-gain system is choked and uniform, the minimum curve $\eta_{\text{min}}$ of the system’s total pressure recovery coefficient is derived theoretically, assuming stoichiometric hydrogen/air mixtures as reactants, as shown in figure~\ref{fig20}. The results indicate that the curves of $\eta_{\text{RDE}}$ and $\eta_{\text{min}}$ follow a similar trend. Specifically, both $\eta_4$ values first increase and then decrease with the rise of $M_1$, peaking at $M_1 = 1$. However, the $\eta_{\text{RDE}}$ curve consistently lies above $\eta_{\text{min}}$. Furthermore, compared to $\eta_{\text{min}}$, $\eta_{\text{RDE}}$ can achieve positive PG over a broader range of $M_1$. This difference is primarily due to $\eta_{\text{RDE}}$ being derived from an actual RDE system, accounting for the non-uniformity of the exit flow. Therefore, for RDEs with stoichiometric hydrogen/air mixtures, this study provides a more general PG criterion that RDEs can achieve positive PG when the inlet Mach number is between $M_{1,cr1}$ and $M_{1,cr2}$.

\section{Conclusion}
\label{sec5}

This study employs an abstract check valve model to simulate flow channel configurations with varying flow characteristics. The model channel remains fully open during forward flow, while it contracts in the case of backflow. The backflow check strength, denoted as $\alpha_b$, is defined such that a larger $\alpha_b$ indicates a stronger blocking effect on backflow. The quasi-1D assumption is adopted along the axial direction to simplify the radial features of the annular RDE, and the quasi-2D RDE flow governing equations are established. Numerical results regarding the variation of PG with different expansion ratios $A_e$ and backflow check strengths $\alpha_b$ are analyzed. The main conclusions are as follows:

Effective suppression of back-propagating pressure can be achieved either by increasing $A_e$ (to accelerate the incoming flow) or by increasing $\alpha_b$ (to reduce the backflow velocity). However, reverse shock wave structures persist even when no backflow occurs. This suggests that the check valve model is not a completely undisturbed control method. While it suppresses back-propagating pressure, it inevitably perturbs the original flow field, and such perturbations cannot be entirely eliminated by increasing the backflow check strength. Moreover, under non-ideal check conditions, the length of the combustor's intake block zone exceeds that of the backflow region due to the backflow of the detonation products.

In the quasi-2D RDE, the exit total pressure recovery coefficient $\eta_4$ is less affected by $\alpha_b$. In contrast, $\eta_4$ increases rapidly as $A_e$ decreases, and positive PG can be achieved when $A_e < 3.336$. For the same $\alpha_b$, to prevent inlet blocking, the propagation of back-propagating pressure to the throat exit defines the minimum $A_e$ under this $\alpha_b$ condition. The larger the $\alpha_b$, the smaller the corresponding minimum $A_e$. There exists a restrictive relationship between $\alpha_b$ and the minimum $A_e$ required to achieve positive PG. Specifically, the minimum $A_e$ can only be lower than 3.336 when $\alpha_b > 51.1$. Therefore, if the suppression effect on back-propagating pressure is insufficient, the quasi-2D RDE cannot achieve positive PG. Enhancing this suppression by optimizing the flow channel configuration represents a core aerodynamic challenge for enabling RDEs to achieve positive PG under current experimental conditions.

The exit total pressure recovery coefficient $\eta_4$ of the quasi-2D RDE is positively correlated with $\eta_a$, the total pressure recovery coefficient induced by the cut-off NSW. When the expansion ratio $A_e$ is small, the intensity of the cut-off NSW is weak. In this case, increasing $\alpha_b$ to avoid inlet blocking becomes crucial for maintaining positive PG.

By normalizing the quasi-2D RDE and adopting the isentropic assumption, this study provides a more general PG criterion for RDEs with stoichiometric hydrogen/air mixtures: positive PG can be achieved when the inlet Mach number is between 0.177 and 2.749.

\section*{Declaration of Competing Interest}
\label{sec:competing-interest} 

The authors declare that they have no known competing financial interests or personal relationships that could have appeared to influence the work reported in this paper. 

\section*{Acknowledgment}
\label{sec:acknowledgment} 

This work is supported by the Natural Science Foundation of China (NSFC, Grant No. 52306152).

\appendix 
\clearpage 

\section{ Existence of the forward flow solution}\label{appA}

When the gas flows forward, the flow channel of the check valve model is fully open ($A_{th} = A_{lb}$). The forward flow check strength $\alpha_f = -1$ is obtained via equation (\ref{a2.6}). Substituting $\alpha_f = -1$ and equation (\ref{a2.9}) into equation (\ref{a2.11}) yields:

\begin{equation}
\Delta s = \frac{R}{\gamma - 1} \ln\left( \frac{\frac{1}{2}x - \frac{\gamma + 1}{2(\gamma - 1)}}{\frac{1}{2} - \frac{\gamma + 1}{2(\gamma - 1)}x} x^\gamma \right)
\label{A1}
\end{equation}

Assuming the forward flow is isentropic, let $\Delta s = 0$. Combining with equation (\ref{a2.9}), we can obtain:

\begin{equation*}
x=1,y=1
\label{A2}
\end{equation*}

\section{ Existence of the backflow solution}\label{appB}

When the backflow occurs, the check valve model exerts a blocking effect on the passing gas ($A_{th} < A_{lb}$). According to equation (\ref{a2.6}), the backflow check strength $\alpha_b > -1$.

Equation (\ref{a2.9}) can be rewritten as follows:

\begin{equation}
y = \frac{g(x)}{f(x)}
\label{B1}
\end{equation}

Where

\begin{equation}
f(x) = \frac{\gamma + 1}{2(\gamma - 1)} x^2 + \alpha_b \frac{\gamma}{\gamma - 1} x + \frac{1}{2}
\label{B2}
\end{equation}

\begin{equation}
g(x) = -\frac{1}{2} x^2 + \frac{\gamma}{\gamma - 1} x + \frac{1}{2} + \alpha_b \frac{\gamma}{\gamma - 1}
\label{B3}
\end{equation}

Let the roots of $f(x) = 0$ be $x_{f1}$ and $x_{f2}$ (where $x_{f1} < x_{f2}$), and the roots of $g(x) = 0$ be $x_{g1}$ and $x_{g2}$ (where $x_{g1} < x_{g2}$). To explore the range of $x$ corresponding to the condition that $M_{lb}$ is real when $\alpha_b$ takes different values, the combination of algebra and geometry is adopted. By plotting the curves of the functions and observing the features of their intersections as well as the magnitudes of the function values, the interval where physical solutions exist is intuitively reflected. Characteristic parameters $\alpha_i$ ($i=1,2,\ldots,5$) are selected as interval division points to partition the interval $(-1,\infty)$. These points are arranged in ascending order as follows:

\begin{equation*}
\alpha_1 = -\sqrt{1 - \frac{1}{\gamma^2}},  \alpha_2 = \frac{1 - \gamma}{2\gamma},  \alpha_3 = 0,  \alpha_4 = \sqrt{1 - \frac{1}{\gamma^2}}, \alpha_5 = 1
\label{Helm}
\end{equation*}

\begin{table}[h!] 
  \centering 
  \setlength{\tabcolsep}{7pt} 
  \caption{Results of the properties analysis of $f(x)$ and $g(x)$.}
  \label{tab3}
  \begin{tabular}{lcc} 
    \toprule 
    Function & $f(x)$ & $g(x)$ \\[3pt]
    \midrule 
    Opening direction & up & down \\
    Axis of symmetry & $x=-\frac{\gamma}{\gamma+1} \alpha_b$ & $x=\frac{\gamma}{\gamma-1}$ \\
    Y-axis intersection & $\left(0,\frac{1}{2}\right)$ & $\left(0,\frac{1}{2}+\alpha_b \frac{\gamma}{\gamma-1}\right)$ \\
    Zero existence & $\makecell{\Delta_f=\left(\alpha_b \frac{\gamma}{\gamma-1}\right)^2 - \\ \frac{\gamma+1}{\gamma-1} > 0}$ & $\makecell{\Delta_g=\left(\frac{\gamma}{\gamma-1}\right)^2 + 1 + \\ \frac{2\alpha_b \gamma}{\gamma-1} > 0}$ \\
    Function intersections & \multicolumn{2}{c}{$\makecell{\left(1,(1+\alpha_b) \frac{\gamma}{\gamma-1}\right), \\ \left(-\alpha_b,\frac{1-\alpha_b^2}{2}\right)}$} \\
    \bottomrule 
  \end{tabular}
\end{table}

\begin{figure}
  \centerline{\includegraphics[width=\textwidth]{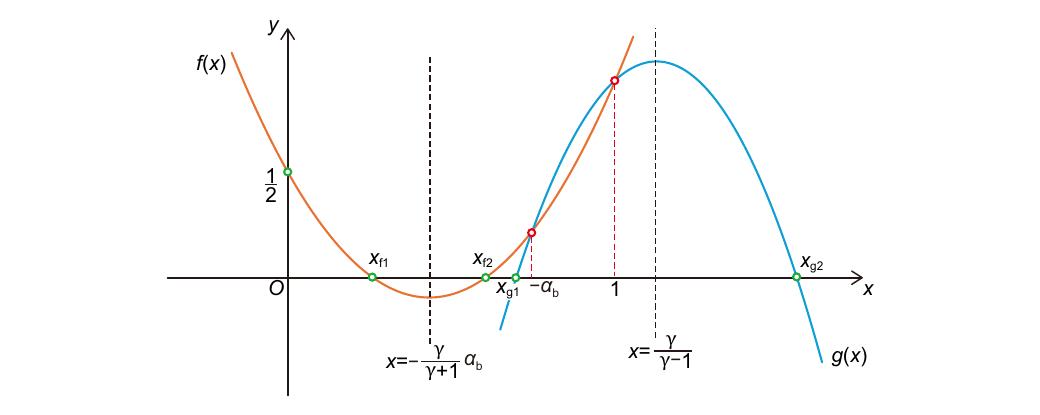}}
  \caption{The relative position between $f(x)$ and $g(x)$.}
\label{fig21}
\end{figure}

Taking the interval of $\alpha_b \in (-1, \alpha_1)$ as an example, the specific process of determining the existence of physical solutions for the check valve model is explained as follows by combining algebra and geometry. First, the properties of $f(x)$ and $g(x)$ are analyzed based on equations (\ref{B2}) and (\ref{B3}), with the results presented in table~\ref{tab3}. Then, the relative position between $f(x)$ and $g(x)$ is plotted in figure~\ref{fig21}. It can be observed that the distribution characteristics of the function zeros and intersections, as well as the magnitude relationships of the function values in different intervals, are clearly illustrated. The existence of physical solutions is as follows:

\begin{enumerate}
    \item When $x \in (0, x_{f1}) \cup (x_{f2}, x_{g1}) \cup (x_{g2}, \infty)$, functions $f(x)$ and $g(x)$ have opposite signs. According to equation (\ref{B1}), this leads to $y < 0$, which contradicts the non-negativity of parameters in actual physical processes. Therefore, there are no physical solutions in these intervals.
    
    \item When $x \in (x_{f1}, x_{f2})$, $x < -\alpha_b$ and $g(x) < f(x) < 0$ hold, resulting in $y > 1$. This result satisfies the constraint $M_{lb}^2 > 0$ in equation (\ref{a2.10}), indicating that this interval is valid for physical solutions.
    
    \item When $x \in (x_{g1}, -\alpha_b)$, $f(x) > g(x) > 0$ applies, yielding $0 < y < 1$. This violates the constraint $M_{lb}^2 > 0$, so no solutions exist in this interval.
    
    \item When $x \in (-\alpha_b, 1)$, $g(x) > f(x) > 0$ holds, leading to $y > 1$. This also fails to satisfy the condition $M_{lb}^2 > 0$, thus no physical solutions exist here.
    
    \item When $x \in (1, x_{g2})$, $x > -\alpha_b$ and $f(x) > g(x) > 0$ are satisfied, resulting in $0 < y < 1$. This meets the constraint $M_{lb}^2 > 0$, making this interval valid for physical solutions.
\end{enumerate}

In summary, physical solutions exist in the interval $x \in (x_{f1}, x_{f2}) \cup (1, x_{g2})$.

For the remaining intervals of \( \alpha_b \), the same analytical method as above is adopted to determine the valid range of the density ratio \( x \) that satisfies both \( y > 0 \) and \( M_{lb}^2 > 0 \), as shown in table~\ref{tab4}. Additionally, this table also systematically organizes the existence of zeros for \( f(x) \) and \( g(x) \), as well as the relative positions between these zeros and the intersections of the two functions.

\begin{table}[h!] 
  \centering 
  \setlength{\tabcolsep}{5pt} 
  \caption{Statistics of the density ratio $x$ (satisfying $y > 0$ and $M_{lb}^2 > 0$), the existence of zeros for $f(x)$ and $g(x)$, and the relative positions between these zeros and the intersections of the two functions, under different intervals of $\alpha_b$.}
  \label{tab4}
  \resizebox{\linewidth}{!}{
    \begin{tabular}{lccccc} 
      \toprule 
      $\alpha_b$ & $\Delta_f$ & $\Delta_g$ & \makecell[c]{Relative positions between zeros\\and intersections} & $x$ & $y$ \\[3pt]
      \midrule 
      $\alpha_b \in (-1, \alpha_1)$ & $\Delta_f>0$ & $\Delta_g>0$ & $0<x_{f1}<x_{f2}<x_{g1}<-\alpha_b<1<x_{g2}$ & $x \in (x_{f1},x_{f2})$ & $y>1$ \\
      & & & & $x \in (1,x_{g2})$ & $0<y<1$ \\
      $\alpha_b \in (\alpha_1, \alpha_2)$ & $\Delta_f<0$ & $\Delta_g>0$ & $0<x_{g1}<-\alpha_b<1<x_{g2}$ & $x \in (1,x_{g2})$ & $0<y<1$ \\
      $\alpha_b \in (\alpha_2, \alpha_3)$ & $\Delta_f<0$ & $\Delta_g>0$ & $x_{g1}<0<-\alpha_b<1<x_{g2}$ & $x \in (1,x_{g2})$ & $0<y<1$ \\
      $\alpha_b \in (\alpha_3, \alpha_4)$ & $\Delta_f<0$ & $\Delta_g>0$ & $x_{g1}<-\alpha_b<0<1<x_{g2}$ & $x \in (1,x_{g2})$ & $0<y<1$ \\
      $\alpha_b \in (\alpha_4, \alpha_5)$ & $\Delta_f>0$ & $\Delta_g>0$ & $x_{g1}<-\alpha_b<x_{f1}<x_{f2}<0<1<x_{g2}$ & $x \in (1,x_{g2})$ & $0<y<1$ \\
      $\alpha_b \in (\alpha_5, \infty)$ & $\Delta_f>0$ & $\Delta_g>0$ & $x_{f1}<-\alpha_b<x_{g1}<x_{f2}<0<1<x_{g2}$ & $x \in (1,x_{g2})$ & $0<y<1$ \\
      \bottomrule 
    \end{tabular}
  }
\end{table}

\clearpage
\bibliographystyle{unsrtnat} 
\bibliography{reference}    

\end{document}